\documentclass[iop,apj]{emulateapj} \usepackage{apjfonts} \newif\ifdraft \draftfalse \newif\ifsub \subfalse \newif\ifpre \pretrue
\usepackage{graphicx}
\usepackage{color}

\shorttitle{The VLBA Galactic plane survey}
\shortauthors{Petrov et al.}
\received{2011 January 6}
\accepted{2011 May 1}
\published{2011 June 22}
\submitted{}
\ifpre
\else
\fi
\bibpunct{(}{)}{,}{a}{}{,}
\makeindex
\citeindextrue
\newcommand{\n}{\nodata}
\newcommand{\getlength}[1]{\ifx#1\end \let\next=\relax
            \else\advance\count255 by1 \let\next=\getlength\fi \next}
\newcount\switch
\newcommand{\Begmat}{\ifm\switch=1\else\switch=0$\fi}
\newcommand{\Endmat}{\ifnum\switch=0$\fi}
\newcommand{\ifnularg}[1]{ \count255=0 \getlength#1\end \ifnum\count255=0 }
\newcommand{\ifm}{\makebox{}\ifmmode}

\long\def\ifundefined#1#2#3{\expandafter\ifx\csname
  #1\endcsname\relax#2\else#3\fi}
\newcommand{\beq}   { \begin{eqnarray} }
\newcommand{\eeq}[1]{ \ifnularg{#1} end{eanarray} \else
                      \label{#1}\end{eqnarray}    \fi }
\newcommand{\eeqn}{\nonumber\end{eqnarray}}
\newcommand{\flo}[2]{ {\Begmat
            \ifnularg{#1} 10^{#2}\else \mbox{#1} \cdot 10^{#2}\fi\Endmat}}

\newcommand{\Frac}[2]{\frac{\displaystyle\strut #1}{\displaystyle\strut #2} }
\newcommand{\ntab}[2]{ \multicolumn{1}{#1}{#2} }
\newcommand{\nntab}[2]{ \multicolumn{2}{#1}{#2} }

\newcommand{\der }[2]{\Frac{ \partial #1 }{ \partial #2 } }
\newcommand{\Sum}{\displaystyle\sum}
\newcommand{\vhex}{ \vspace{0.5ex} }

\newcommand{\hm}{\hphantom{-}}
\newcommand{\hp}{\hphantom{+}}
\newcommand{\ha}{\hphantom{a}}
\newcommand{\dss}{\displaystyle}
\newcommand{\vex}{\vspace{1ex}}
\newcommand{\lp}{ \left(  }
\newcommand{\rp}{ \right) }
\newcommand{\ds}{\displaystyle}
\newcommand{\eff}{\rm eff}
\newcommand{\Tr}{\rm Tr}
\newcommand{\Cov}{\rm Cov}
\newcommand{\bra}[1]{\!\!<\!\!#1\!\!>}
\newcommand{\tra}[1]{\!{{#1}^{\!\bf \top}}\! }
\newcommand{\PIMA}{$\cal P\hspace{-0.067em}I\hspace{-0.067em}M\hspace{-0.067em}A$ }

\newcommand{\Number}[1]{\ifnum#1<10\relax0\number#1\else\number#1\fi}
\newcommand{\isodate}{
\count151=\time
\divide\count151 by 60
\count151=\count151
\multiply\count151 by 60
\count152=\time
\advance\count152 by -\count151
\divide\count151 by 60
\count152=\count151
\multiply\count151 by 60
\count153=\time
\advance\count153 by -\count151
\Number{\year}.\Number{\month}.\Number{\day}--\Number{\count152}:\Number{\count153}
}
\definecolor{Dred}{rgb}{0.312,0.070,0.070}
\definecolor{Dblue}{rgb}{0.070,0.070,0.312}
\definecolor{Dgreen}{rgb}{0.070,0.312,0.070}
\definecolor{Db}{rgb}    {0.050,0.0,0.320}

\ifpre
\fi
\newcounter{note}
\let\oldmarginpar\marginpar
\renewcommand\marginpar[1]{\-\oldmarginpar[\raggedleft\footnotesize #1]%
{\raggedright\footnotesize #1}}
\newcommand{\Note}[1]{\Rdb{#1}{\addtocounter{note}{1}%
\marginpar{\small\underline{\Rdb{Corr \arabic{note}}}}}}
\newcommand{\note}[1]{\Rdb{#1}}
\newcommand{\YNote}[1]{\Rdb{#1}{\addtocounter{note}{1}%
\marginpar{\small\underline{\Rdb{Corr \arabic{note}}}}}}
\newcommand{\ynote}[1]{\Rdb{#1}}
\renewcommand{\note}[1]{#1}
\renewcommand{\Note}[1]{#1}
\renewcommand{\ynote}[1]{#1}
\renewcommand{\YNote}[1]{#1}

\begin{document}

\title{The VLBA Galactic Plane Survey --- VGaPS }

\author{L. Petrov}
\affil{ADNET Systems Inc./NASA GSFC, Greenbelt, MD 20771, USA}
\email{Leonid.Petrov@lpetrov.net}
\author{Y. Y. Kovalev}
\affil{Astro Space Center of Lebedev Physical Institute,
       Profsoyuznaya 84/32, 117997 Moscow, Russia}
\affil{National Radio Astronomy Observatory,
       P.O.~Box 2, Green Bank, WV 24944, USA}
\email{yyk@asc.rssi.ru}
\author{E. B. Fomalont}
\affil{National Radio Astronomy Observatory,
       520 Edgemont Road, Charlottesville, VA~22903--2475, USA}
\email{efomalon@nrao.edu}
\author{D. Gordon}
\affil{NVI Inc.,/NASA GSFC, Code 698.2 Greenbelt, MD 20771, USA}
\email{David.Gordon-1@nasa.gov}

\begin{abstract}


This paper presents accurate absolute positions from a 24~GHz Very
Long Baseline Array (VLBA) search for compact extragalactic sources in
an area where the density of known calibrators with precise
coordinates is low. The goals were to identify additional sources
suitable for use as phase calibrators for galactic sources, determine
their precise positions, and produce radio images.  In order to
achieve these goals, we developed a new software package, \PIMA, for
determining group delays from wide-band data with much lower detection
limit. With the use of \PIMA\ we have detected 327 sources
out of 487 targets observed in three 24 hour VLBA experiments.  Among
the 327 detected objects, 176 are within 10 degrees of the Galactic
plane. This VGaPS catalogue of source positions, plots of
correlated flux density versus projected baseline length, contour plots,
as well as weighted CLEAN images and calibrated visibility data in FITS
format, \note{are} available on the Web at \url{http://astrogeo.org/vgaps}.
Approximately one half of objects from the 24~GHz catalogue were observed
at dual band 8.6~GHz and 2.3~GHz experiments. Position differences at
24~GHz versus 8.6/2.3~GHz for all but two objects on average are strictly
within reported uncertainties. We found that for two objects with
complex structure positions at different frequencies correspond
to different components of a source.

\end{abstract}


\section{Introduction}
\label{s:introduction}

The method of very long baseline interferometry (VLBI), first proposed
by \citet{r:mat65}, has numerous applications in the areas of high
resolution imaging, differential astrometry, absolute astrometry,
space geodesy, and space navigation. Because the turbulence in the
neutral atmosphere and ionospheric fluctuations \Note{set} a limit of
coherent averaging at typically 1 to 10 minutes, depending on
frequency, the detection of weak sources \Note{that require longer
integration} is not possible.

To overcome this limitation, the majority of VLBI observations are
made in phase referencing mode: the telescopes of the array slew
rapidly between a weak target and a nearby strong calibrator. The
phase changes of the calibrator trace the fluctuations in the
atmosphere, and when they are subtracted from the phase of the target,
the residual phases are essentially free from fluctuations caused by
the atmosphere, and the target integration time can be extended
almost indefinitely, enabling detection and imaging of weak objects.
This technique is called phase-referencing.

The technique of phase referencing \Note{allows also to determine}
the precise differential position of a target with respect to
a calibrator with accuracy reaching $0.05$~mas or better, even with moderate
signal-to-noise ratio (SNR).  The advantage of differential astrometry
over absolute astrometry is that the contribution of unaccounted
propagation delays and errors in station positions is diluted by
a factor of the target-to-source separation in radians. Either
the target or calibrator may be observed in a narrow band.

The ability to image weak sources and determine their positions
accurately with respect to a nearby calibrator have made phase
referencing very popular. According to \citet{r:wrobel_pr}, in
2003--2008, 63\% of VLBA observations used this technique.  However,
to make phase referencing possible, a dense catalogue of phase
calibrators is needed such that a suitable calibrator will be found
within 2--$3^\circ$ of any target.  And for precise differential
astrometry, the calibrator position accuracy of a few milliarcsecond
is needed.  Efforts to create such a catalogue of calibrators
commenced in the 1980s under the NASA's Crustal Dynamic
Project program \citep{r:rya87} that ultimately resulted in the ICRF
catalogue of 608 sources \citep{r:icrf98}. Later, over \Note{6000 sources}
were observed in the framework of the VLBA Calibrator Survey (VCS)
program \citep{r:vcs1,r:vcs2,r:vcs3,r:vcs4,r:vcs5,r:vcs6}, the VLBA
RDV program \citep{r:rdv}, and the continuing Australian Long Baseline
Array Calibrator Survey (LCS) program \citep{r:lcs1}. By the end of
2010, the number of known calibrators with position accuracies better
than 5~mas surpassed 4600. The probability of finding a calibrator within
$2^\circ$ of any direction is currently 64\%, and within $3^\circ$ is
90\%.  However, the distribution of calibrators is not uniform on the
sky due to several factors: the large scale structure of the Universe;
the location of most observing stations in the northern
hemisphere; and obscuration and confusion within $5^\circ$ of the
Galactic plane.

Finding calibrators in the Galactic plane region is especially difficult
for several reasons. First, the region is filled with many galactic
objects, and surveys from single antennas or km-sized arrays, which
are needed to find calibrator candidates, avoid this region; for
example JVAS \citep{r:jvas} and AT20G \citep{r:at20g}.  Therefore, the
pool of available candidates is limited near the plane.  Second, many
potential candidates with flat spectra are extended galactic objects,
such as planetary nebulae or HII regions, that cannot be detected by
VLBI.  Finally, the apparent angular size of extragalactic objects
observed through high plasma density near the Galactic plane are
broadened by Galactic scattering, and cannot be detected \Note{at low
frequencies} on baselines longer than several thousand kilometers.

Nevertheless, a dense grid of calibrators in the Galactic plane is
needed for studies of compact galactic objects in both continuum
emission (pulsars, X-ray binaries) and line emission (egs. water
masers, methanol masers, Hydrogen absorption). \Note{Some extragalactic
sources near the Galactic plane have been associated
with \textit{Fermi}-detected $\gamma$-ray objects
\citep{r:1FGL,r:1LAC}} only through VLBI calibrator surveys as
it was suggested and successfully shown by \cite{r:Kovalev2009}.

In order to increase the density of calibrators in the Galactic plane
and to search for suitable calibrators within $2^\circ$ of known water
masers, we developed an observing strategy that combine the VERA and
VLBA arrays.  First, we systematically screened 2462 sources with
declinations $>-40^\circ$ using the 4-station VLBI array VERA
\citep{r:VERA} at
22~GHz (K-band) by observing each object in two 120-second long
scans. We detected 533 objects, 180 of them new at K-band, and these
results are given in \citet{r:vera_fss}.  Since \Note{precise
determination of parallaxes and proper motions} of sources with water-maser
emission \Note{is} one of the main targets of the VERA project, potential
calibrators near known water masers were preferentially \Note{included}
in the observations.

This paper describes the follow-up VLBA observations at 24 GHz of
\Note{487 radio sources} in order to determine their precise
positions and images.  We denote these observations as the VLBA
Galactic Plane Survey (VGaPS), and \Note{the results of this observing
campaign are described} in this paper. VLBA observations, source selection
and the scheduling algorithm are given in section \S\ref{s:observations}.
The data analysis procedure is presented in \S\ref{s:anal}.
For analysis of these observations we have developed a new approach
for wide-band fringe search and astrometric analysis which is described
in detail in \S\ref{s:PIMA}.  The validation of the results made with
the new approach is given in \S\ref{s:validation}. The images are
described in section \S\ref{s:images}, the source position catalog
is listed in \S\ref{s:catalogue}, and the K- and S/X-band astrometric
VLBI positions are compared in \S\ref{s:KSXcomparison}.  The results are
summarized in section \S\ref{s:summary}.

\section{Source selection and observations}
\label{s:observations}

The VLBA observations were made at 24~GHz for several reasons. First,
Galactic scattering at low frequencies broadens images and degrades
source positions.  Even at the frequency of 8.6 GHz, many calibrators
near the Galactic plane are too broadened to be useful.  Second, many
galactic radio targets are associated with H${}_2$O maser emission
at 22.5~GHz.  Since the correlated flux
density of calibrators generally decreases with increasing frequency,
a good calibrator at 2.3~and 8.6~GHz may not be sufficiently compact
or bright for use at 22~GHz.  Third, source positions at 22~GHz may
not necessarily be the same as those at 8.6~GHz because of the effects
of frequency dependent source structure, especially for
multi-component objects, and because of \YNote{frequency dependent core
shifts \citep[e.g.,][]{r:lob98,r:kov08,r:sokol11}.}

  We considered the task of extending the pool of calibrators for
Galactic astrometry more broadly: not only to increase the list of compact
extragalactic radio sources within $10^\circ$ of the Galactic plane,
but also to re-observe known sources at K-band that are either in the Galactic
plane or close to known masers at higher galactic latitudes. Water masers
are the main target of the VERA project, and determination of their
parallaxes and proper motions is important for studying the three-dimensional
structure and dynamics of the Galaxy's disk and bulge, and for revealing
the true shape of the bulge and spiral arms, its precise rotation curve and the
distribution of dark matter. Dual-beam K-band VERA observations require
calibrators with positions known to the milliarcsec level within
$2.\!^{\circ}2$ of the target.

\subsection{Source selection}
\label{s:selection}

   \Note{VLBI can detect emission only from a compact component of
a source, which should be bright enough to be detected above the
noise background. Chances to find a radio source sufficiently compact
for VLBI detection are} significantly enhanced if information about source
spectra is available.  For a majority of sources with flat spectrum,
\Note{i.e. with the spectral index $\alpha > -0.5 \enskip (S\sim f^{\alpha})$},
synchrotron emission from a compact core often dominates, so such sources
are often compact.  In regions more than $10^\circ$ from the Galactic plane,
comparison of surveys at different frequencies can be used to select
the sources with flat spectra.  However, in the crowded Galactic
plane, source misidentification and confusion from surveys at
different frequencies often result in ambiguous information about a
source spectrum.  Also, there are many extended sources with thermal
emission, such as planetary nebulae and HII regions, that have flat
spectrum but are \Note{too extended} for VLBI detection.

  To determine if a candidate source was sufficiently strong and compact
to be a VLBI calibrator, we first used the VERA array at 22~GHz
\citep{r:vera_fss}.  In the survey $\sim$1400 objects within $6^\circ$
of the Galactic plane and $\sim$1000 objects within $2^\circ$ of known
maser sources outside the Galactic plane were observed with the
four-element VERA array at baselines of 1000--2000~km long in two
scans each, and approximately 20\% were detected. \Note{Among the
533 detected sources, 305 objects are with galactic latitude
$|b|<10^\circ$, and 228 objects are with $|b|>10^\circ$. The list of
detected sources included all objects with the probability of
false detection $< 0.1$.} These objects formed \Note{the first} set
of candidate calibrators.

  We also added 239 objects in the Galactic plane and 44 objects outside
Galactic plane, selected on the basis of their spectra using the
Astrophysical CATalogs support System CATS \citep{r:cats}.  This
database currently includes catalogues from \Note{395} radio astronomy
surveys. We selected all entries with sources within a $20''$-radius
and with measurements of flux density at least \Note{at} three frequencies
in the range of 1.4--100~GHz.  We fit a straight line to the logarithm
of the spectrum and then estimated the spectral index and the flux
density extrapolated to 24~GHz.  We selected 451 objects with
extrapolated correlated flux densities at K-band \Note{$>$ 200~mJy}, spectral
index flatter than $-0.5$ and $|b|<10^\circ$. \Note{Then, we scrutinized
this list and removed sources within $30''$ of known planetary
nebulae or HII regions and sources with anomalous spectra that indicated a
possible misidentification.} \YNote{The remaining list was added to the
pool of calibrators. This list contained also Sagittarius A$^*$ since
it has never before been observed with VLBI in the absolute astrometry
mode.}

This set of sources formed the pool of 816 candidate objects to be
\Note{followed-up with the VLBA}.  Because of the large number
of candidate objects, a priority from 1 to 4 was assigned to each source.
The first priority was given \Note{to 180 new sources with correlated
flux densities in the range 100--300~mJy detected with VERA that have
never been observed with VLBI before}. The second was given to
sources in the Galactic plane detected with VERA; the third was given
to other sources detected with VERA; and the least priority was given
to sources outside the Galactic plane not detected with VERA. All new
sources detected with VERA were scheduled.  Source outside of the
Galactic plane were included for logistical reasons. Since the
Galactic plane has an inclination with respect to the equator of
$62^\circ\!.6\,$, the density of candidate sources near the Galactic
plane is a non-uniform function of right ascensions. We thus included
sources outside the Galactic plane in the right ascension regions that
had significantly fewer candidate sources than others in order to
avoid gaps in the schedules.

In addition to the target sources, we also selected 56 objects from
the K/Q survey \citep{r:kq} to serve as amplitude and atmosphere
calibrators. These are the brightest sources at K-band with precisely
known positions and publicly available images in FITS-format.

\subsection{Scheduling algorithm}
\label{s:scheduling}

   \Note{A sufficiently good preliminary VLBA observing schedule for three
24-hour sessions was prepared automatically. Then it was manually
adjusted in order to produce a more efficient final schedule.}

%
   The three sets of information needed in order to
compile the schedule were: 1)~the list of the 816 target sources with
their J2000 positions and priority level; 2)~the list of the 56
calibration sources at 24 GHz; and 3)~the two or three sidereal times
ranges (at the array center at station {\sc pietown}) that each source
should be observed.  These times were a function of the source
declination.  For example, a source with $\delta>50^\circ$ could be
observed three times with very flexible time ranges; whereas, a source
with $\delta<-35^\circ$, must have tight ranges in order to be
observed by at least 8~VLBA antennas at an elevation higher than ten
degrees.

\Note{The scheduling goal was to observe each source for three scans of
120 seconds length, unless it was south of declination $-25\degr$,
in which case only two scans were scheduled.}  With an average overhead
of about 45~sec between sources, about 1550~scans over the three days
(72~hours) could be scheduled.  Every 90~minutes, four scans were reserved
for atmosphere calibrators. The algorithm then began filling in observing
slots, taking the highest priority sources first, and those at the lowest
declinations, since these have minimal scheduling flexibility.
The algorithm scheduled sources that were relatively close in the sky
in order to minimize slew times, which can be as long as three minutes.

The slots containing the calibrators were chosen in
order to maximize the range of elevations for the VLBA antennas.
In practice, this meant maximizing elevations of observed sources at
{\sc sc-vlba, mk-vlba} and {\sc pietown}.  It was important to have one
low elevation observation for all telescopes, and this could be
scheduled by observing a source either in the far north and/or in the
far south.  All three days were scheduled at the same time, since it
did not matter on which day any of the three sidereal time slots
occurred for a source.

The final schedule optimization was done 'by hand' and consisted of
several steps.  First, some sources could be placed in only one or two
slots and would have to be removed if another slot could not be
found. Since about 10\% of the slots could not be filled using the
automatic algorithm because the source list was not uniformly distributed
over the sky, additional slots were usually found to complete a
source's schedule requirement.  This often meant bending some of the
rules or moving calibration sources or blocks by five to fifteen
minutes.  In regions where the target source density was large, scan
integrations were decreased from 120~sec to 110~sec.  All sources with
priority 1 were included.

The next stage insured that the calibration scans were optimized in
order to obtain good elevation coverage for the antennas.  The purpose
of these observations was 1)~to serve as amplitude and bandpass
calibrators; 2)~to improve robustness of estimates of the path delay
in the neutral atmosphere; and 3)~to tie the source positions of new
sources to existing catalogues, such as the ICRF catalogue
\citep{r:icrf98}. The final stage of optimization tweaked the
observing schedule in obvious ways in order to save slewing time since
the scheduling algorithm did not minimize slew times, and because of
the above adjustments to insure proper observing coverage for each
source.  The schedule for each session was then carefully checked
using the NRAO SCHED program to insure that each scan had sufficient
integration time, and no more than one of the antennas was below ten
degree elevation (except for sources south of $-35^\circ$).

The results of the scheduling and ultimate detection for each priority
group is given in Table~\ref{t:det}.  The number of target sources
selected in each group is given in column 2, those scheduled in column
3, those detected by the VLBA in column 4, and the detection rate in
the last column.  The first two groups are sources detected in VERA
Fringe Search Survey, the last two groups contain other objects. The
table does not include the 56 sources used as atmosphere calibrators.

\begin{table}
   \caption{VLBA Target Sources from VERA Observations}
   \label{t:det}
   \centering
   \begin{tabular}{lrrrr}
       \hline\hline
       group                 &  pool & scheduled & detected & ratio \\
       \hline
       Galactic, VERA        &  305  &       184 &      140 & 76\%  \\
       Non-galactic, VERA    &  228  &       151 &      133 & 88\%  \\
       Galactic  others      &  239  &       108 &       36 & 33\%  \\
       Non-galactic, others  &   44  &        44 &       15 & 34\%  \\
       \hline
       Total                 &  816  &       487 &      327 & 67\%  \\
   \end{tabular}
\end{table}

\subsection{Observations}
\label{s:data}

The VGaPS observations were carried out in three 24 hour observing
sessions at the VLBA on 2006 June 04, 2006 June 11, and 2006 October
20.  \Note{Each target source was observed in several scans: 3 sources
in 1 scan, 356 sources in 2 scans, 1 source in 3 scans, 124 in 4 scans
and 3 sources in 5 scans. The scan durations were 100--120
seconds. In total, antennas spent 57\% of time on target sources.}
In addition to these target objects, 56 strong sources
previously observed at K-band were taken from the astrometric and
geodetic catalogue \mbox{\tt 2004f\_astro}%
\footnote{\tt \url{http://astrogeo.org/rfc}}.

\begin{table}
   \caption{The lower edge IF frequencies (in GHz)}
   \label{t:frq}
   \centering
   \begin{tabular}{ll}
      \hline\hline
      IF1  & 24.20957 \\
      IF2  & 24.22257 \\
      IF3  & 24.26157 \\
      IF4  & 24.33957 \\
      IF5  & 24.48257 \\
      IF6  & 24.58657 \\
      IF7  & 24.65157 \\
      IF8  & 24.67757 \\
      \hline
   \end{tabular}
\end{table}

The data digitized at four levels were recorded with the rate of
256~Mbit/s in eight 8~MHz wide intermediate frequencies (IF) bands
spread over a bandwidth of 476~MHz (Table~\ref{t:frq}). The
frequencies were selected to minimize the amplitude of sidelobes of
the Fourier transform of the bandpass.

\section{Data analysis}
\label{s:anal}

The data were correlated in Socorro on the VLBA hardware correlator.
The correlator output contains the complex spectra of the autocorrelation
function and the spectrum of the cross-correlation function for each
accumulation period. The  accumulation period was chosen to be 0.131072~s,
and the spectral resolution was set to 0.5~MHz, i.e. 16 spectral channels
per IF.

After correlation, the data were stored in a file compliant with the
FITS-IDI specifications (Eric Greisen, NRAO memo
N114\footnote{Available at {\small\tt \url{ftp://ftp.aoc.nrao.edu/pub/software/\\
aips/TEXT/PUBL/AIPSMEM114b.PS}}.}).  Small a~priori amplitude
corrections were applied; bad data were flagged (sometimes bad data
are not found until later processing); and the definitions of the
reference frequencies were modified.

Further data analysis involves computation of group delays for
each scan and each baseline, computation of theoretical path delay, and
then fitting parameters of the linear model into the differences
between the observed and theoretical delays using least squares
(LSQ).

\subsection{Traditional narrow-band fringe fitting algorithm}

 The data set then contains 128 data streams for each of the 45
baselines if all ten VLBA antennas are operating: 8 IF's, each with 16
equally-space frequency channels.  After the complex bandpass
calibration, the phase difference among all of the frequency streams
remains unchanged, and the relevant residual phase parameters
associated with any scan are 1)~the residual phase at the scan midpoint,
2)~the average group delay (phase gradient with frequency), and
3)~the average delay rate (delay gradient with time), for each antenna.
These parameters are called the {\it residual phase
terms}. These parameters are estimated with a fringe fitting procedure.
Generally, one antenna is chosen as the reference ---
a well-behaved one near the array center --- so that set of a residual
phase terms are associated for each scan and all other antennas.
These parameters are functions of many astrometric quantities (source
position, site positions, antenna-based tropospheric path delays,
Earth Orientation parameters, etc), which can be determined from
analysis systems like Calc/Solve from data obtained from carefully
prepared observing schedules. Source structure and dispersive phase
effects (e.g. ionosphere) produce a non-linear phase/frequency
relationship.

The algorithm that determines the residual phase terms is
\Note{implemented in the AIPS task FRING, and in the past, all VLBA
observations under the absolute astrometry and geodesy programs, such as
VCS, RDV, and K/Q, were processed with use of this
software \citep{r:aips}. When IFs are spread over a wide band with gaps,}
the AIPS algorithm determines the residual phase in two steps: first,
for each of the 8 IF's the residual phase, single-band group delay and
rates are independently obtained. Then, the residual phases of each IF are
fit to a linear phase versus frequency term to produce the group delay,
using AIPS program MBDLY. This procedure is described in detail
in \citet{r:rdv}.

\subsection{Wide-band baseline-based fringe fitting algorithm}
\label{s:PIMA}

  The two-step approach has a substantial shortcoming: it
requires fringe detection for each individual IF independently
within a narrow band. \Note{Using all $N$ IFs simultaneously for
coherent averaging, we can detected a source with the amplitude $\sqrt{N}$
smaller than using only one IF. This degradation of the detection limit does
not pose a problem for most geodesy observations, or for absolute astrometry
of bright sources, since the target SNR is usually very high, but it
impacts significantly absolute astrometric experiments of weak
sources. The traditional AIPS algorithm did not detect a sizeable
fraction of the sources observed in the VGaPS experiment.}

This limitation motivated us to develop an advanced algorithm for
wide-band fringe fitting across all of the IF's within the band.
For logistical reasons, instead of augmenting AIPS with the new
task, we decided to develop from scratches the new software package
called \PIMA\footnote{Available at {\tt \url{http://astrogeo.org/pima}}.}
that is supposed to replace the AIPS for processing absolute astronomy
and geodesy experiments. Here we outline the method of wide-band fringe
search used for processing this experiment.

\subsubsection{Spectrum re-normalization}
\label{s:renorm}

   Digitization of the input signal and its processing with a digital
correlator causes an amplitude distortion with respect to an ideal analogue
system. As documented by \citet{r:kog_autocorr}, many effects distort
both cross-correlation and auto-correlation spectrum exactly the same way.
Therefore, if we divide the cross-correlation spectrum by the auto-correlation
spectrum averaged over time and over frequencies within each IF, we will remove
these distortions. However, there are two effects that affect cross-
and auto-spectrum differently because the amplitude of the cross-spectrum
is very low, and the amplitude of the auto-correlation spectrum is close to 1.

   The non-linear amplitude distortion of the digitized signal was studied
in depth by \citet{r:kog_ampl} who derived a general expression for the
correlation coefficient of the digitized signal as a function of the
correlation coefficient of the hypothetical analogue signal. In the absence
of fringe stopping, the output correlation coefficient $\rho_{out}$ is
expressed via the correlation coefficient for an analogue case $\rho$ as
\beq
   \begin{array}{lcl}
     \rho_{out} &=& \hp  2\, \kappa \, \dss \int\limits_0^\rho \Frac{1}{\sqrt{1 - \rho^2}} d \rho
                 \\ & &
                 + \: 2 \,\kappa \,(n-1) \, \dss \int\limits_0^\rho \Frac{1}{\sqrt{1 - \rho^2}}
                     \lp e^{-\frac{v^2}{2 (1-\rho^2)}} +
                         e^{-\frac{v^2_1}{2 (1-\rho^2)}}
                     \rp d \rho
                 \\ & &
                + \: \kappa \,(n-1)^2 \, \dss \int\limits_0^\rho \Frac{1}{\sqrt{1 - \rho^2}}
                     \lp e^{-\frac{v^2_1 - 2\rho v_1 v_2 + v^2_2}
                           {2 (1-\rho^2)}} +
                         e^{-\frac{v^2_1 + 2\rho v_1 v_2 + v^2_2}
                           {2 (1-\rho^2)}}
                     \rp d \rho ,
   \end{array}
\eeq{e:e1}
  where $\kappa$ is the normalization coefficient, $n$ is the ratio of two
level of quantization, $v_1$ and $v_2$ are the dimensionless digitizer
levels in units of variance of the input signal. Their numerical values
compiled from the paper of \citet{r:kog_bfact} are presented in
table~\ref{t:levels}.

\begin{table}[hb]
   \caption{\rm Numerical coefficients in integral \ref{e:e1} for three cases
                of the number of bits per sample: (1,1), (1,2), (2,2)}
   \centering
   \begin{tabular}{l @{\qquad} l @{\qquad} l @{\qquad} l @{\qquad} l}
        \hline\hline
              & $n$    & $v_1$  & $v_2$  & $\kappa$ \\
        \hline
        (1,1) & 1.0    & 0.0    & 0.0    & 0.3803   \\
        (1,2) & 3.3359 & 0.0    & 0.9816 & 0.05415  \\
        (2,2) & 3.3359 & 0.9816 & 0.9816 & 0.07394  \\
        \hline
   \end{tabular}
   \label{t:levels}
\end{table}

  \Note{When $\rho \ll 1$, the dependence $\rho_{out}(\rho)$ becomes linear:
$\rho_{out} = 2/\pi \cdot \rho \approx 0.6366\, \rho $ for the case of 1-bit
sampling,} and $\rho_{out} \approx 0.8825\, \rho $ for the case of 2-bit
sampling. Therefore, distortion of the cross-spectrum is proportional to
$\rho_{out}/\rho$, and dividing the cross-spectrum by 0.6366 or 0.8825 we
eliminate the distortion of the fringe amplitude introduced by
the digitization.

  However, the amplitude of the autocorrelation spectrum is close to 1,
and we cannot ignore non-linearity of $\rho_{out}(\rho)$. \Note{To correct
the autocorrelation spectrum for digitization distortion,} we follow the
procedure outlined by \citet{r:kog_autocorr}. First, the auto-correlation is
inverse Fourier transformed. It should be noted that the correlator provides
the autocorrelation for $N$ spectral channels for non-negative frequencies
from 0 to $N-1$. We restore the autocorrelation at the $N$-th channel
by linear extrapolation using the $N-2$ -th and $N-1$ -th values of the
spectrum and set the spectrum for $N-1$ negative frequencies to zero.
The dimension of the Fourier transform is $2N$. Second, the result of the
transformation, the auto-correlation coefficient versus time lag,
is de-tapered, i.e. divided by the self-convolution of the weighting
function. The VLBA correlator normally uses uniform weighting, i.e.
uses weight 1 for all points. The self-convolution of the uniform
weighting is a triangle function $\wedge$(i):
\beq
   \wedge (i) = \left\{
     \begin{array}{l @{\qquad} l}
         \frac{1}{N}( N - |i| ) & \mbox{if} \quad |i| < N \vex \\
         0                      & \mbox{otherwise}
     \end{array}
   \right.
\eeqn

  Third, the correlation function is divided by its maximum that is found
at the zero lag. Fourth, each correlation coefficient is divided by
$\rho_{out}(\rho)$. Fifth, the correlation function is multiplied back by
the stored value at the zero lag. Sixth, the correlation function is again
tapered by multiplying it by $\wedge$(i). Finally, we perform the Fourier
transform of the corrected correlation function and get the re-normalized
auto-spectrum, free from digitization distortion. The square root of the
product of the auto-correlation spectra from two stations of a baseline,
averaged over time and frequency within each IF, gives us an estimate of
the fringe amplitude scale factor. But before dividing the
cross-correlation spectrum by this scale-factor we have to take into
account a specific effect of the hardware VLBA correlator\footnote{The new
VLBA software DiFX correlator does not have this problem.}. An insufficient
number of bits in internal correlator registers resulted in a decrease of
the amplitude when its is large, \Note{and when it happens,
the autocorrelation spectra are corrupted.} \citet{r:kog_bfact} suggested
the following model for accounting for this effect:
\beq
   F = 1 + \Frac{w}{4 \, S \, A \, V_s} ,
\eeq{e:e2}
  where $w$ is the weight of the spectrum data equal to the ratio of processed
samples to the total number of samples in an accumulation period, $A$
is the accumulation period length, $S$ is 2 when the correlator processed
single polarization data and 1 if both polarizations were correlated, and
$V_s$ is the visibility scale factor provided by the correlator. We divide
the autocorrelation spectrum by the $F$ factor.

\subsubsection{Coarse fringe search}
\label{s:coarse}

  The correlator output provides autocorrelation and cross-correlation
spectra for each pair of baselines and each scan. The cross-correlation
spectrum is computed at a uniform two-dimensional grid of accumulation
periods and frequencies and is accompanied with weights that are the ratio
of the number of processed samples in each accumulation period to the
nominal number of samples.

  The fringe fitting procedure searches for phase delay $\tau_p$,
phase delay rate $\dot{\tau}_p$, group delay $\tau_g$, and its
time derivative $\dot{\tau}_g$ that correct their a~priori values used
by the correlator model in such a way that the coherent sum of
weighted complex cross-correlation samples \Note{$c_{kj}$
\beq
   \begin{array}{ll}
      C(\tau_p,\tau_g,\dot{\tau}_p,\dot{\tau_g}) = &
          \dss \sum_k \sum_j c_{kj} \, w_{kj} \times \\ &
          \hspace{-3em}
          e^{i( \omega_0 \tau_p \; + \;
                \omega_0  \dot{\tau}_p (t_k-t_0) \; + \;
                (\omega_j - \omega_0) \tau_g \; + \;
                (\omega_j - \omega_0) \dot{\tau}_g (t_k - t_0) )}
     \vex
   \end{array}
\eeq{e:e3}
  reaches the maximum amplitude. Index $k$ runs over time, and
index $j$ runs over frequencies. $\omega_0$ and $t_0$ denote angular reference
frequency within the band and the reference time within a scan and $w_{kj}$}
are weights. Function $C(\tau_p,\tau_g,\dot{\tau}_p,\dot{\tau_g})$ is
essentially non-linear, and we need a really good starting value in order
to find the global maximum by traditional optimization algorithms. We can
notice that term $2\pi \, \omega_0 \tau_p$ in expression~\ref{e:e3} does
not depend on the summation indexes, and $\dot{\tau}_g$ is usually small.
Therefore, for the purpose of coarse fringe search we simplify
expression~\ref{e:e3} to
\beq
   \hspace{-1em}
     C(\tau_p,\tau_g,\dot{\tau}_p) e^{-i2\pi \, \omega_0 \tau_p} \approx &
          \dss \sum_k \sum_j c_{kj} \times
                e^{i( \omega_0  \dot{\tau}_p (t_k-t_0) \; + \;
                      (\omega_j - \omega_0) \tau_g )}.
\eeq{e:e4}

  For the search of the maximum, the trial functions
$C(\tau_p,\tau_g,\dot{\tau}_p)$ are computed on a dense grid of the search
space $\tau_g,\dot{\tau}_p$. It follows immediately from expression~\ref{e:e4}
that $|C| = |{\cal{F}}(c_{kj})|\,$,
where $\cal{F}$ denotes the two dimensional Fourier transform.

  The first step of the coarse fringe search is to compute the two-dimensional
Fast Fourier Transform (FFT) of the matrix of the cross-correlation
spectrum. The first dimension of the matrix runs over time, the second
dimension runs over frequency. The sampling intervals are $\Delta t/\beta$
and $\Delta f/\gamma$ where $\Delta t$ and $\Delta f$ are the duration
of the accumulation period and the spectral resolution respectively, and
$\beta$ and $\gamma$ are integer oversampling factors. The elements of the
matrix which do not have measurements or have discarded measurements are
padded with zeroes. The dimensions of the matrix are chosen to have the power
of 2 for gaining the maximum performance of the FFT numerical algorithm.

  The oversampling factors greater than 1 are used for mitigation of amplitude
losses. The FFT produces the estimates of |C| at a discrete grid of
group delays and delay rates. If the maximum of the amplitude of the
coherent sum of the cross correlation function samples falls at group delays
and delay rates between the nodes of the grid, its magnitude will be greater
than the amplitude of |C| at the nearest grid point by a factor $L$.

  The amplitude loss factor $L$ at the coarse search matrix is the
integral average over \Note{time and frequency}:
\beq
   \begin{array}{ll}
      L = & \Frac{1}{t_s} \dss \int^{t_s/2}_{-t_s/2}
            \cos 2\pi \lp \omega_0 \tau_p - \Frac{k}{\beta t_s} \rp
            t \, dt \times \\ &
          \Frac{1}{f_b}   \dss \int^{f_b/2}_{-f_b/2}
            \cos 2\pi \lp \omega_0 \tau_g - \Frac{l}{\gamma f_b} \rp f \,
            d\hspace{-0.05em}f ,
   \end{array}
\eeq{e:e5}
  where $k$ and $l$ are indexes of the nearest grid nodes, $t_s$ is the scan
duration, $f_b$ is the total bandwidth. The integral \ref{e:e5} is easily
evaluated analytically, and the maximum losses
$L = \mbox{sinc}\lp\pi/\lp2\beta\rp\rp \cdot \mbox{sinc}\lp\pi/\lp2\gamma\rp\rp$
are achieved when $\dot{\tau_p}$ and $\tau_g$ happen to be just between grid
nodes. In the case when the oversampling factor is 1,
$L = 4/\pi^2 \approx 0.405$. That loss factor effectively raises the detection
limit by $1/L = 2.467$ in the worst case. We used the grid
$4096 \times 4096$ \Note{which corresponds to $\beta=4, \gamma=4$ for
$t_s = 120$~s, $f_b = 476$~MHz. Therefore, the
maximum loss factor for our experiment is 0.959,} which results in raising
the detection limit by no more than $4.1\%$.

  The group delay and delay rate that correspond to the maximum of
the discrete Fourier transform of the cross-correlation matrix provide the
coarse estimates of group delay and delay rate. Their accuracy depends on
the grid resolution. The next step is to refine their estimates.
The first stage of the fine search is an iterative procedure that
computes |C| at a progressively finer three-dimensional grid at the close
vicinity of the maximum using the discrete 3D Fourier transform. The third
dimension is group delay rate, omitted during the coarse search. Dimensions
of the transform for group delay, phase delay rate, group delay rate are
3, 3, and 9 respectively. At the first step of iterations, the grid runs
over $\pm 1$ element of the coarse grid for group delays and phase delay
rates and in the range $\pm 2 \cdot 10^{-11}$ for group delay rate. After
each step of iterations, the grid centered around the maximum element
shrinks its step by 2. In total, 8 iterations are run. The phase of the
coherent sum of cross correlation function determined with the iterative
procedure according to expression~\ref{e:e4} is the fringe phase with
the opposite sign.

\subsubsection{The probability of false detection}
\label{s:prob}

The significance of the fringe amplitude depends on
the level of noise. In the absence of signal, the amplitude of the cross
correlation function $a$ has the Rayleigh distribution:
\beq
    p(a) = \frac{a}{\sigma^2} e^{-\frac{a^2}{2\sigma^2}} ,
\eeq{e:e6}
  where $\sigma$ is the standard deviation of the real and imaginary part of
the cross correlation function and $n$ is the total number of spectrum points.
In the case when all points of the spectrum are {\it statistically independent},
the cumulative distribution function of the coherent sum over $n$ points
is \citep{r:tms}
\beq
    P(a) = \left( 1 - e^{-\lp\frac{ a^2}{2\sigma^2}\rp} \right)^n .
\eeq{e:e7}

  Differentiating this expression, we find the probability density function
of the ratio of the amplitude of the coherent sum of spectrum samples
of the noise to its standard deviation as
\beq
    p(a) = n \Frac{a}{\sigma^2} e^{-\frac{a^2}{2\sigma^2}}
           \left( 1 - e^{-\frac{a^2}{2 \sigma^2}}\right)^{n-1} .
\eeq{e:e8}

  Under an assumption that all samples are {\it statistically independent},
the variance of the noise of the coherent sum of $N$ samples is scaled as
$\sigma = \sigma_s/\sqrt{t_s\, S_r}$, where $S_r$ is the sampling
rate of the recorded signal ($6.4 \cdot 10^{7}$ in our experiment),
$t_s$ is the scan duration, and is $\sigma_s$ the variance of an
individual sample , 1 for a perfect system.

  However, the assumption of statistical independence is an idealization.
The presence of systematic phase errors, deviation of the bandpass from the
rectangular shape and other factors distort the distribution. The deviation
from the statistical independence is difficult to assess analytically.

  We evaluate the variance of the noise by estimating the variance over
a sample of 32768 random points at the region of the Fourier transform
of the coherent sum of the cross-correlation that does not contain the signal.
The indexes of grid elements of the sample are produced by using the
random number generator. The sample of amplitudes is ordered, and one half
of the points greater than the median is rejected. The variance over
remaining points is computed and an iterative procedure is launched that adds
back previously rejected points in the ascending order of their amplitudes
and it updates the mean value and variance. The iterations are run till
the maximum amplitude of the next sample reaches $3.5\sigma_a$. The initial
rejection and consecutive restoration of points with amplitudes
$> 3.5\sigma_a$ ensures that no points with the signal from the source
affect computation of $\sigma_a$ and $\bra{a}$. The rejection of the tail
of the amplitude distribution causes a bias in estimate of the mean, but
the magnitude of the bias is only $-2 \cdot 10^{-4}$, which is negligible.
We define a signal to noise ratio as SNR = $a/\bra{a}$. It follows immediately
from expression \ref{e:e6} that $\: \bra{a} = \sqrt{\frac{\pi}{2}} \sigma$.

We assume that the a~posteriori distribution of the signal to noise ratios
$s=a/\sigma_a$ can be approximated as a function like
expression~\ref{e:e8} with effective parameters $\sigma_{\eff}$ and $n_{\eff}$:

\beq
    p(s) = \Frac{2}{\pi} \Frac{n_{\eff}}{\sigma_{\eff}} \, s \,
           e^{-\frac{s^2}{\pi}}
           \left( 1 - e^{-\frac{ s^2}{\pi}} \right)^{n_{\eff}-1} .
\eeq{e:e9}

  These parameters $\sigma_{\eff}$ and $n_{\eff}$ can be found by fitting
the left tail of the empirical distribution of the signal to noise ratios.
Using their estimates, we can evaluate the probability of finding the
amplitude less than $a$ if no signal is present, i.e. the probability
of false detection:
\beq
    P_f(s) = 1 - \ds\int\limits_0^s p(s) \, ds =
             1 - \frac{1}{\sigma_{\eff}}
                 \left( 1 - e^{-\frac{s^2}{\pi}} \right)^{n_{\eff}}.
\eeq{e:e10}

\begin{figure}[t!]
  \centering
  \ifpre
     \includegraphics[width=0.48\textwidth,clip]{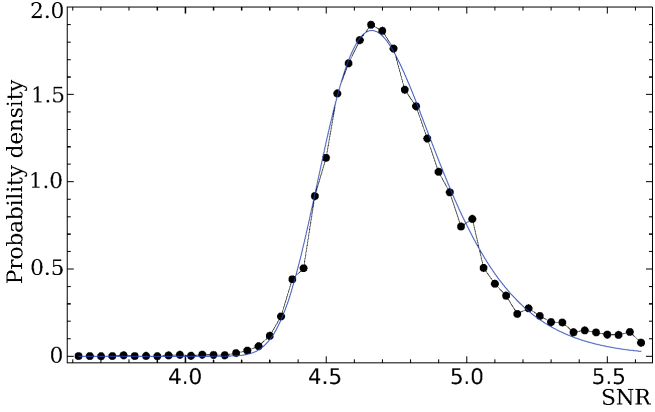}
  \else
     \includegraphics[width=0.48\textwidth,clip]{snr_distr_1200.eps}
  \fi
  \caption{\note{The low end of the empirical distribution of the achieved
           SNR for the fringe amplitude from results of fringe fitting
           VLBA data (filled circles) and the fitted curve (thin line)
           of the theoretical distribution for the case of no signal.}
  }
  \label{f:snr_distr}
\end{figure}

  The low end portion of the empirical SNR distribution and the table of
the probability of false detection for VGaPS experiments are shown in
Figure~\ref{f:snr_distr} and Table~\ref{t:prob_false}.

\begin{table}[bth]
   \caption{\rm The probability of false detection as a function of SNR}
   \centering
      \begin{tabular}{l @{\qquad\quad} l}
            \hline\hline
            SNR  & $ P_f(s) $   \\
            \hline
            4.96 & 0.3          \\
            5.19 & 0.1          \\
            5.61 & 0.01         \\
            5.99 & 0.001        \\
            6.68 & $10^{-5}$    \\
            \hline
      \end{tabular}
   \label{t:prob_false}
\end{table}

\subsubsection{Fine fringe search}
\label {s:fine}

  Finally, the group delay, phase delay rate, group delay rate and fringe
phase at the reference frequency are estimated using LSQ
in the vicinity of the maximum provided its amplitude exceeds the detection
limit. The goal of the LSQ refinement is to get realistic estimates of
statistical errors of fitted parameters and to account for possible systematic
errors by analyzing residuals. All cross-correlation spectrum data points
of a given observation are used in a single LSQ solution with weights
reciprocal to their variance.

  Determining the variance of the fringe phase of an individual point is
trivial only in two extreme cases: when \mbox{SNR $\ll$ 1}, and when
\mbox{SNR $\gg$ 1}. In the first case the fringe phase distribution
becomes uniform in the range of $[0, 2\pi]$ with the variance
$\pi/\sqrt{3} \approx 1.813$. In the second case expanding the expression
for fringe phase in the presence of noise
$\phi = \arctan\frac{S_i + n_i}{S_r + n_r}$, where $S_i$ and $S_r$ are the real
and imaginary parts of the signal and $n_i$ and $n_r$ are the real and
imaginary parts of the noise, into the Taylor series, neglecting
terms O($n/S^2$), and evaluating the variance of the expansion, we get
$\sigma_\phi = \sqrt{\frac{2}{\pi}}\frac{1}{\mbox{\sc snr}}$. For the general
case, the problem becomes more difficult, since the $\sigma_\phi$ depends
on the variance of the noise and on the amplitude of the signal non-linearly.
An analytical solution requires evaluation of complicated integrals that are
not expressed via elementary functions.

\begin{figure}[t!]
  \centering
  \ifpre
     \includegraphics[width=0.48\textwidth,clip]{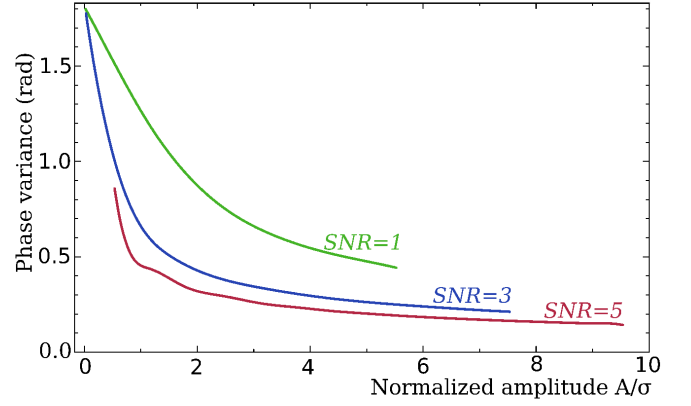}
   \else
     \includegraphics[width=0.48\textwidth,clip]{snr_sigma_1200.eps}
  \fi
  \caption{The variance of fringe phase as a function of the normalized
           amplitude $A/\sigma_n$ in the presence of a signal with given
           SNR.
  \label{f:snr_sigma}
  }
\end{figure}

   We used the Monte Carlo approach for computing these variances. Let us
consider a random complex set $s = A + n_r + i n_i$, where $A$ is the
amplitude of the simulated signal and $n_r, n_i$ are independent random
variables with Gaussian distribution. Their variance $\sigma$ was
selected in such a way that $ A = \sqrt{\frac{\pi}{2}} \sigma \, \mbox{SNR}$.
Then for a given
SNR, we can compute the variance of phase of $s$ as a function of the
normalized amplitude $|s|/\sigma$. This is done by generation a long series
(1 billion points) of the simulated complex signal for a given SNR,
computing the amplitude and phase of the time series, splitting the signal
into a uniform grid of 128 bins over normalized amplitude that spans
the interval [SNR $-$ 4.5, SNR $+$ 4.5], computing the variance of the phase
of the simulated signal over all points that fall into \Note{each bin}, and
approximating the dependence of $\sigma_\phi(|s|/\sigma)$ with a smoothing
B-spline of the 3rd order over 6 nodes. We computed $\sigma_\phi(|s|/\sigma)$
for SNRs in the range [0, 12.7] with steps of 0.1. Examples of this dependence
for several SNRs are shown in figure~\ref{f:snr_sigma}. The set of B-spline
coefficients forms a two dimensional table with axes SNR and normalized
amplitude that allows evaluating $\sigma_\phi$ for a given SNR and a given
fringe amplitude.

\begin{figure}[ht]
  \centering
  \ifpre
      \includegraphics[width=0.48\textwidth,clip]{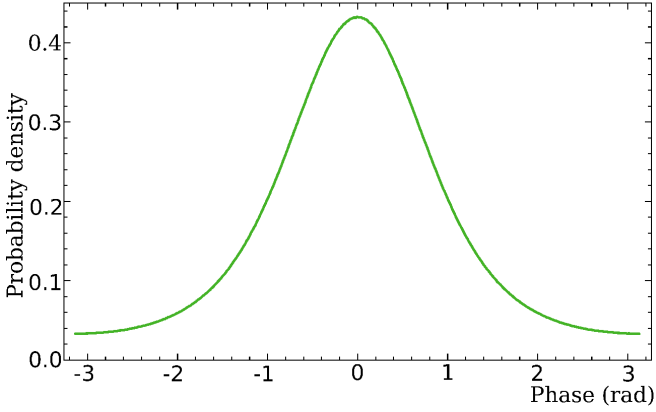}
  \else
      \includegraphics[width=0.48\textwidth,clip]{phs_dist_snr_1_1200.eps}
  \fi
  \caption{The probability density distribution of fringe phase with SNR=1.
  \label{f:snr_1}
  }
\end{figure}

  It should be noted that the SNR of the coarse search is related to the
amplitude \Note{coherently averaged over {\it all} valid cross-correlation
spectrum samples, $ \sum_k^{N_t}\sum_j^{N_s} w_{kj} $, where $N_t$ is
the number of accumulation
periods and $N_s$ is the total number of spectral channels. The SNR of
{\it an individual} accumulation period (the elementary SNR) is
$ \sqrt{ \sum_k^{N_t}\sum_j^{N_s} w_{kj} } $ times less}. For our
experiments, the typical reduction of the SNR is a factor of 340. This means
that for almost all sources the elementary SNR will be less than 1.
As we have seen previously, the distribution of fringe phases at very low
SNR's is close to uniform with a variance of $\pi/\sqrt{3}$. The phase
becomes uncertain due to the $2\pi$ ambiguity, and the LSQ estimation
technique loses its diagnostic power. Therefore, the cross-correlation
function with applied phase, phase
delay rate and group delays computed during the coarse fringe search have
to be coherently averaged over time and frequency in segments large enough
to have sufficiently high SNR over a segment to provide an ambiguous phase.
As it is seen in Figure~\ref{f:snr_1}, the fringe phase distribution at SNR=1
is already sharp enough for that. Therefore, the number of spectral channels
and the number of accumulation periods within a segment is chosen in such
a way that the SNR be at least 1. Marginally detected scans with SNR=5 have
24 segments that average all spectral channels within an IF and over
1/3 of the scan interval.

  Using all segments, we determine four fitting parameters $\bf p$ using
the LSQ:
\beq
    {\bf p} = (\tra{A} \, {\cal W} \, A)^{-1}
     \tra{A} \, {\cal W} \mbox{\boldmath $\phi_s$} ,
\eeq{e:e11}
  where $A$ is the matrix of observations, ${\cal W}$ is the diagonal weight
matrix and {\boldmath $\phi_s$} is the vector phases of the
cross-correlation function averaged over segments.

  The mathematical expectation of the square of the weighted sum of residuals
$R$ in the presence of noise $\epsilon$ is
\beq
    \hspace{-1em}
    E(R) \, = \, \Tr\lp {\cal W} \Cov \epsilon \rp \; - \;
           \Tr\lp \Cov{\epsilon} \,
             {\cal W} A (\tra{A} {\cal W} \, A)^{-1} \tra{A} \,
             {\cal W} \rp ,
\eeq{e:e12}
  which is reduced to $n - m$  where $n$ is the number of equations and $m$
is the number of estimated parameters if the weight matrix ${\cal W}$ is
chosen to be $(\Cov \epsilon){}^{-1}$.

  The presence of additive errors, for example fluctuations in the atmosphere,
will increase  $\Cov \epsilon$ and our estimate of the error variance based on
the amplitude of the spectrum sample without knowledge of the scatter
of the cross-spectrum phases is incomplete. A small additive noise with
the variance $k$ times less than the amplitude of the signal affects the
amplitude as $O(k^2)$, but it affects phase as $O(k)$. One of the measures
of the incompleteness of the error model is the ratio of the square
of the weighted sum of residuals to its mathematical expectation.

  We can extend our error model assuming that the weight matrix is
${\cal W} = {\Cov \epsilon}^{-1} - q^2 I$, where $I$ is the unit matrix
of the same dimensions as ${\cal W}$ and $q$ is the parameter. This model
is equivalent to an assumption that the used squares of weight of each segment
were less than the true one by some parameter $q$ equal to all segments.
The mathematical expectation of $R$ for this additive weighting model is
\beq
    E(R) = n - m \: + q^2 \left[
           \Tr\lp {\cal W} \rp \: - \:
           \Tr\lp {\cal W} (A {\cal W}^{2} \, A)^{-1} \tra{A} \rp
           \right].
\eeq{e:e13}
  Inverting \ref{e:e13}, we find $q$ for a given $E(R)$:
\beq
    \begin{array}{ll}
    q = \sqrt{ \Frac{E(R) \: - \: (n - m)}
               { \Tr\lp {\cal W} \rp \: - \:
                 \Tr\lp {\cal W} (A {\cal W}^2 \, A)^{-1} \tra{A} \rp}.
             }
    \end{array}
\eeq{e:e14}

  Replacing the mathematical expectation of $R$ with its value evaluated
from the residuals, we can find the re-weighting parameter $q$ for a given
solution. Several iterations provides a quick convergence of $\Frac{E(R)}{R}$
to 1. Applying an additive reweighting constant results in an increase in
estimates of parameter uncertainties. It may happen that $q$ becomes imaginary.
This means that ${\cal W}$ was overestimated. In our analysis we set $q=0$
in that happens.

\subsubsection{Complex bandpass calibration}

  Coherent averaging of the cross-correlation spectrum over frequencies
assumes that the data acquisition system does not introduce a distortion
of the recorded signal, but this is usually not the case. Each intermediate
frequency has its own arbitrary phase offset and group delay that may
vary with time. The imperfection of baseband filters results in a
non-rectangular shape of the amplitude response.

  For calibrating these effects, \Note{a rail of narrow-band phase calibrator
signals with spacing of 1~MHz was injected near the receivers.} Two tones
per IF are extracted by the data acquisition hardware, and their phase
and amplitude are available for data analysis. When the phase of the phase
calibration signal is subtracted from fringe phases, the result is referred
to the point of injection of the phase calibration signal and this procedure
is supposed to take into account any phase changes that occurred in the
signal passing through the data acquisition terminal. However, we should be
aware of several complications that emerge when we try to use the benefits
of the phase calibration signal. First, the phases of two tones of the
phase calibration signal separated by 6~MHz, as in our sessions, have
ambiguities. Since the instrumental group delay may reach several phase
turns over a 8~MHz IF, the second phase calibration tone is useless without
resolving the ambiguity. Second, the phase calibration itself may have
a phase offset or may become unstable if its amplitude is not carefully
tuned. Therefore, we need to re-calibrate the phase calibration signal
itself in order to successfully apply it to data.

  We compute the complex bandpass function for each station, except the
reference station, that describes the residual instrumental complex bandpass
after applying the phase calibration signal. The cross-correlation spectrum
needs to be divided by the complex bandpass in order to correct the
instrumental frequency-dependent delay and fading of the amplitude.
The procedure for evaluating the complex bandpass has several steps.

  First, all data are processed applying the first tone of the phase
calibration, i.e. the phase of the phase calibration signal is subtracted from
each phase of the cross-correlation signal at a given IF.

  Second, the bandpass reference station is chosen. Then for each station,
we find an observation \Note{at a baseline} with the reference station that
provided the maximum SNR during the first run. Then for each IF we average
the residual spectrum over time and \Note{perform a linear fit to the residual
phases} for determination of the instrumental group delay. Using this
instrumental group delay, we extrapolate the phase of the phase calibration
signal of the first phase calibration tone to the frequency of the second
phase calibration tone and resolve its phase ambiguity. After that, we re-run
the fringe search procedure for these scans by applying the phase calibration
phase to each IF in the form of
a linear function of phase versus frequency that is computed from two phase
calibration tones with resolved phase ambiguity. The result of the new fringe
search gives a new residual spectrum. Then we averaged the residual
spectrum, over time and over $M$ segments within each IF ($M=2$ in our
experiment). The amplitude of the spectrum is normalized by dividing the
average amplitude over all IFs. The phase and the amplitude of the residual
averaged spectrum are approximated with a Legendre polynomial
of degree 5. The result of this approximation as a complex function of
frequency defines the so-called initial complex bandpass. \Note{Analysis of
residuals of rejected observations helps to diagnose malfunction of
the equipment. For example, one or more video-converters may fail,
which may result in a loss of coherence. In that case, the part of the
affected cross-correlation spectrum is masked out.}

  Third, we refine the complex bandpass. We select $N$ more observations
with the highest SNR from the first run at all baselines ($N$=16 in our
experiment) and re-run the fringe fitting procedure with applied phase
(but not amplitude!) of the initial bandpass. We compute a residual spectrum
averaged over time and $M$ segments for each processed observation and
normalize its amplitude. Then we fit a set of 6 coefficients of the
Legendre polynomial for each station, except the reference one, for each IF,
for both amplitude and phase to the phase and amplitude of the residual
spectrum using a single LSQ solution. Then the residuals are computed
and the observations with the maximum absolute value of residual phases
and residual amplitudes are found separately. If the maximum absolute
value of residual phase or residual amplitude exceeds the predefined limit,
the affecting observation is removed and the solution is repeated.
Iterations are run until either the absolute values of all remaining
residuals are less than the predefined limit or the number of observations
at a given baseline drops below $N/2$. The fitted Legendre polynomial
coefficients are added to the coefficients of the initial bandpass and
the result defines the so-called fine complex bandpass.

\begin{figure*}[tbh]
  \centering
  \ifpre
      \includegraphics[width=0.48\textwidth,clip]{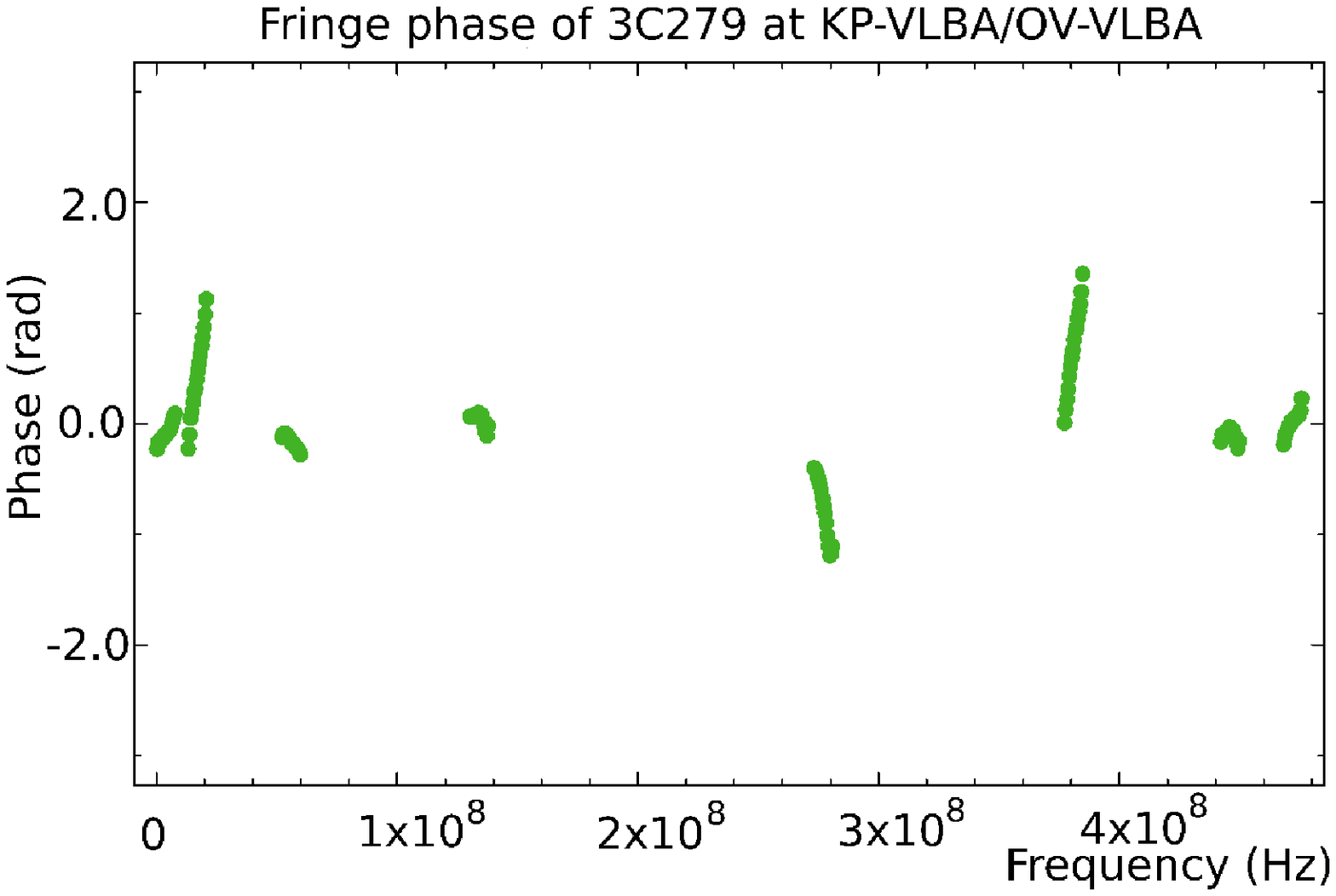}
  \else
      \includegraphics[width=0.48\textwidth,clip]{phas_ovkp_01_1200.eps}
  \fi
  \hspace{0.02\textwidth}
  \ifpre
     \includegraphics[width=0.48\textwidth,clip]{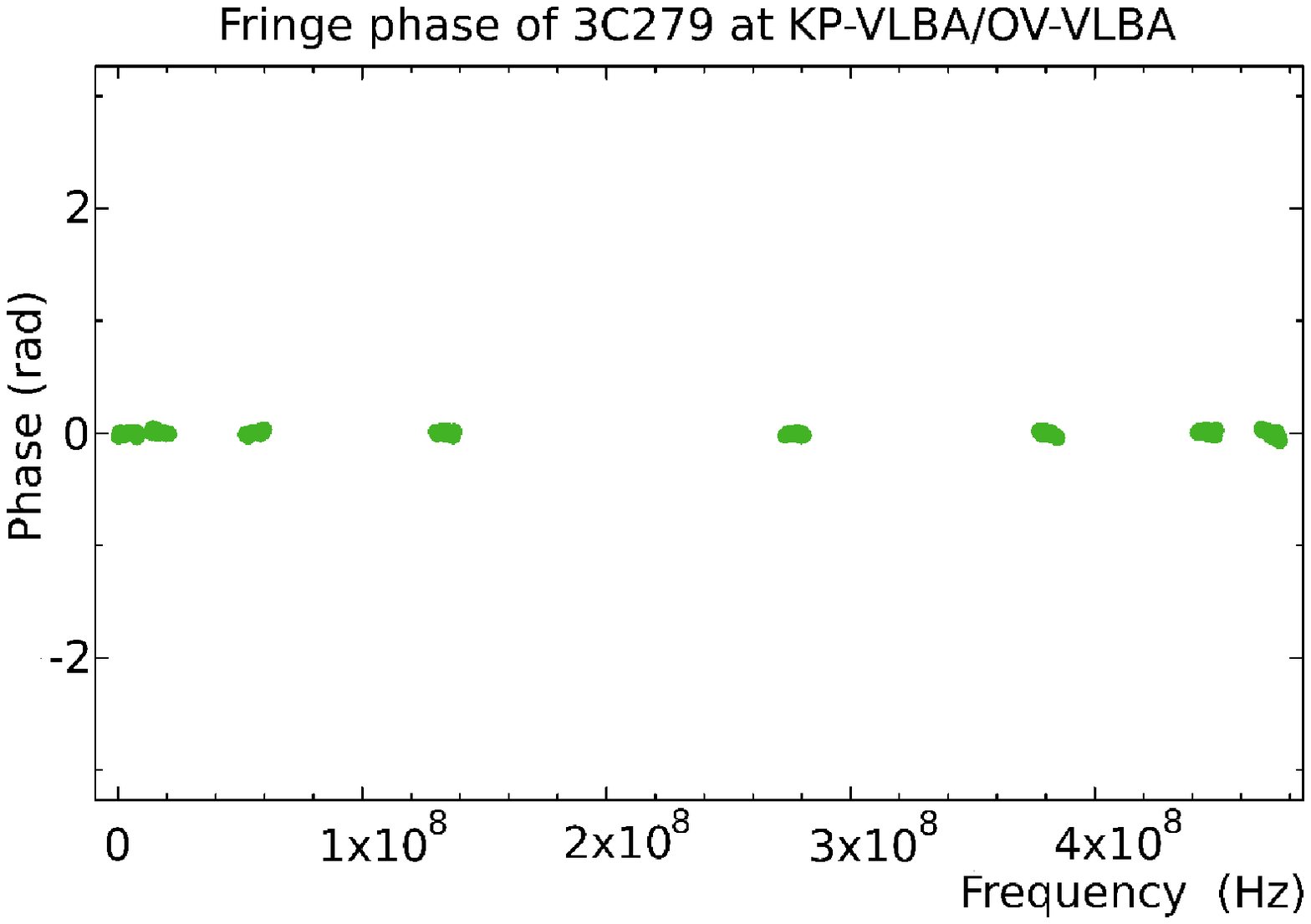}
  \else
     \includegraphics[width=0.48\textwidth,clip]{phas_ovkp_02_1200.eps}
  \fi
  \caption{Residual fringe phases (radians) before (left) and after
           (right) applying bandpass calibration in experiment bp125b.
  \label{f:phcal}
  }
\end{figure*}

  Fourth, all observations are reprocessed with the fine bandpass applied:
the phase of the fine complex bandpass of the remote station of a baseline
is subtracted and the phase of the bandpass of the reference station of
a baseline is added before fringe fitting, and the amplitude is divided
by the square root of the products of bandpass amplitudes after
fringe fitting. Examples of residual phases before and after calibration
are shown in figure~\ref{f:phcal}.

\subsection{Computation of total group delays and phase delay rates}

  The results of the fringe search are residual phase and group delay
as well as their time derivatives determined from analysis of an observation
at a given baseline of a given scan with respect to the a~priori delay
model used by the correlator. We need to compute the total path delay
related to a certain moment of time common to all observations of a scan,
called a scan reference time ($t_{\rm srt}$).

  For logistical reasons, fringe searching at different baselines is performed
independently and therefore, each observation has its own reference epoch,
called a fringe reference times ($t_{\rm frt}$). This time epoch is computed
as the weighted mean epoch of a given observation. Since stations usually
start and end at slightly different time and the number of processed samples
may be different, in general, $t_{\rm frt}$ is different at different
baselines of the same scan. The scan duration may be significantly different
at different stations either by a schedule design when antennas with different
sensitivities participate in observations --- this is often made in geodetic
observations, or due to losses of  some of the observing time at stations for
various reasons. If we set $t_{\rm srt}$ as an average of $t_{\rm frt}$ over
all baselines of a scan, it may happen that for some baselines the difference
$t_{\rm srt} - t_{\rm frt}$ may reach several hundred seconds. Setting a
common $t_{\rm srt}$ for \Note{as many baselines} as possible is desirable since
it allows computation of delay triangle misclosures and some other important
statistics. On the other hand, the uncertainty of the group delay estimate
is minimal at $t_{\rm frt}$. The uncertainty in group delay
at $t_{\rm srt}$ grows as
\beq
   \hspace{-1em}
   \sigma^2_{\tau}(t_{\rm srt}) = \sigma^2_{\tau}(t_{\rm frt})
            \: + \:
            2 \, \Cov( \tau, \dot{\tau}) \, (t_{\rm srt} - t_{\rm frt})
            \: + \:
            \sigma^2_{\dot{\tau}}(t_{\rm srt} - t_{\rm frt})^2,
\eeq{e:e15}
  which is undesirable. In our work, we set the tolerance limit for the growth
of the the uncertainty due to the differences between $t_{\rm srt}$ and
$t_{\rm frt}$: $0.1\sigma$ or 5~ps, whichever is less. Setting this limit,
we find for each observation the maximum allowed $|t_{\rm srt} - t_{\rm frt}|$
by solving quadratic equation \ref{e:e15}. In the case where all observations
of a scan have overlapping intervals of acceptable scan reference times,
we set it to the value that minimizes
$2 \sum \Cov( \tau, \dot{\tau})\, (t_{\rm srt} - t_{\rm frt}) +
 \sigma^2_{\dot{\tau}}(t_{\rm srt} - t_{\rm frt})^2$ over all observations.
In the case where there are observations that have intervals of acceptable
$t_{srt}$ which are not overlapping, the set of observations of a scan
is split into several subsets with overlapping acceptable $t_{\rm srt}$
and the optimization procedure is performed under each subset. Finally,
$t_{srt}$ is rounded to the nearest integer second.

  The VLBA correlator shifts the time tag of data streams from each station
to the moment of time when the wavefront reaches the center of the
coordinate system. This operation facilitates correlation and allows
station-based processing. The a~priori path delay is computed for this
modified time tag. This shift of the time tag depends on the a~priori
parameters of the geometric models, and therefore the total path delay
produced from such a modified quantity would depend on errors of the
a~priori model which would considerably complicate data analysis.
Therefore, we have to undo this shift of the time tag for the reference
station of a baseline for further processing.

  \Note{The correlator delay model for the VLBA hardware and software correlators
is computed as a sum of a 5th degree polynomial fit to the geometric delays,
the linear clock offset, and the coarse  atmospheric model delay over intervals
of 120~s length. We find the a~priori delay of the baseline
reference station $\tau^{\rm rf}_a$ related to the time tag at TAI of the
wavefront arrival to its phase center from the implicit equation}
\beq
  \tau^{\rm rf}_a = \dss\sum\limits_{k=0}^{k=5} a^{\rm rf}_i
                    \lp t_{\rm srt} - t_o -
                        \lp   \tau^{\rm rf}_a
                            - \tau^{\rm rf}_{\rm cl}
                            - \dot{\tau}^{\rm rf}_{\rm cl} \, \tau^{\rm rf}_a
                            - \tau^{\rm rf}_{\rm atm}
                        \rp
                    \rp^k,
\eeq{e:e16}
  \Note{which is solved by iterations. Here $t_o$ is the TAI time tag of
the polynomial start time, $\tau^{\rm rf}_{\rm cl}$ and
$\tau^{\rm rf}_{\rm atm}$ are the clock model and the atmosphere
contribution of the a~priori path model.} We set $\tau^{\rm rf}_a$
on the right hand side of expression \ref{e:e16} to zero for the first
iteration. Three iterations are sufficient to reach the accuracy 0.1~ps.
Using the value $\tau^{\rm rf}_a$ found at the last iteration, we compute the
a~priori path delay for the remote station of the baseline:
\beq
  \tau^{\rm rm}_a = \dss\sum\limits_{k=0}^{k=5} a^{\rm rm}_i
                    \lp t_{\rm srt} - t_o -
                        \lp   \tau^{\rm rf}_a
                            - \tau^{\rm rm}_{\rm cl}
                            - \dot{\tau}^{\rm rm}_{\rm cl} \, \tau^{\rm rf}_a
                            - \tau^{\rm rm}_{\rm atm}
                        \rp
                    \rp^k .
\eeq{e:e17}

  The a~priori delay rate is computed using an expression similar
to equation \ref{e:e17}. Finally, we compute the total path delay
by extrapolating the residual delay to the scan reference time:
\beq
   \tau_{tot} = \tau^{\rm rm}_{\rm apr} - \tau^{\rm rf}_{\rm apr} \: +  \:
                \tau_{\rm res} \: + \:
                \dot{\tau}_{\rm res} (t_{\rm srt} - t_{\rm frt}) .
\eeq{e:e18}

  The delay produced this way is the difference between the interval of
proper time measured by the clock of the remote station between events of
\Note{arrival of} the wave front to the reference point of the remote
antenna and clock synchronization in TAI and the interval of proper time
measured by the clock of the reference station between events
of \Note{arrival of} the wave front to the reference point of the reference
antenna and clock synchronization in TAI.

\subsection{Astrometric Analysis: Delay Modeling}

  Our computation of theoretical time delays in general follows the approach
presented in detail by \citet{r:masterfit} with some refinements. The most
significant ones are the following. The advanced expression for time delay
derived by \citet{r:Kop99} in the framework of general relativity was used.
The displacements caused by the Earth's tides were computed using the numerical
values of the generalized Love numbers presented by \citet{r:mat01}
following a rigorous algorithm described at \citet{r:harpos} with
truncation at a level of 0.05~mm. The displacements caused by ocean loading
were computed by convolving the Greens' functions with ocean tide models
using the NLOADF algorithm of \citet{r:spotl2}. The GOT00 model \cite{r:got99}
of diurnal and semi-diurnal ocean tides, the NAO99 model of ocean zonal
tides \citep{r:nao99}, the equilibrium model \citep{r:harpos} of the pole tide,
and the tide with period of 18.6 years were used. Atmospheric pressure loading
was computed by convolving the Greens' functions with the output of the
atmosphere NCEP Reanalysis numerical model \citep{r:ncep}. The algorithm of
computations is described in full details in \citet{r:aplo}. The empirical
model of harmonic variations in the Earth orientation parameters
{\tt heo\_20091201}\footnote{Available at {\tt\url{http://astrogeo.org/erm}}}
derived from VLBI observations according to the method proposed
by \citet{r:erm} was used. The time series of UT1 and polar motion
from the Goddard operational VLBI solutions were used as a~priori.

  The a~priori path delays in the neutral atmosphere in directions towards
observed sources were computed by numerical integration of differential
equations of wave propagation through the heterogeneous media.
The four-dimensional field of the refractivity index distribution was
computed using the atmospheric pressure, air temperature and specific humidity
taken from the output of the Modern Era Retrospective-Analysis for Research
and Applications (MERRA) \citep{r:merra}. That model presents the atmospheric
parameters at a grid $1/2\degr \times 2/3\degr \times 6^h$ at 72 pressure
levels.

  For considering the contribution of the ionosphere to the
phase of the cross-correlation spectrum, notice that the electromagnetic
wave propagates in a plasma with phase velocity
\beq
    v_p = \Frac{c}
          {\sqrt{ 1 - \Frac{N_v \, e^2}{m_e \, \epsilon_o \, \omega^2} }} ,
\eeq{e:e19}
where $N_v$ --- electron density, $e$ --- charge of an electron, $m_e$ ---
mass of an electron, $\epsilon_o$ --- permittivity of free space, $\omega$ ---
angular frequency of the wave and c --- velocity of light in vacuum. Phase
velocity in the ionosphere is faster than the velocity of light in vacuum.

  After integration along the ray path and expanding expression \ref{e:e19},
withholding only the term of the first order, we get the following expression for
additional phase rotation caused by the ionosphere:
\beq
    \Delta\phi = - \Frac{a}{\omega} ,
\eeq{e:e20}
where $\omega$ is the angular frequency and $ a $ is
\beq
     a = \Frac{e^2}{ 2 \, c \, m_e \,  \epsilon_o }
              \lp \int N_v \, d s_1 - \int N_v \, d s_2 \rp .
\eeq{e:e21}

  Here $s_1$ and $s_2$ are the paths of wave propagation from the source to
the first and second station of the baseline. If $\int N_v \, d s $ is
expressed in units of \flo{1}{16} electrons/$m^2$ (so-called TEC units),
then after having substituted values of constants, we get
$a = \flo{5.308018}{10} \:\: \mbox{sec}^{-1}$ times the difference in the
TEC values at the two stations.

  Since the ionosphere contribution is frequency-dependent, it distorts
the fringe-fitting result. Taking into account that the bandwidth of the
recorded signal is small with respect to the observed frequency, we
can linearize eqn.~\ref{e:e20} near the reference frequency $\omega_o$:
$ \phi = -a/\omega_o + a (\omega_i - \omega)/\omega_o^2$. Comparing it with
expression \ref{e:e4}, we see that the first frequency-independent term
contributes to the phase delay and the second term, linear with frequency,
contributes to the group delay. The fine fringe search is equivalent
to solving the LSQ for $\tau_p$ and $\tau_g$ using
the following system of equations:
\beq
     \tau_p \omega_o + \tau_g (\omega_k - \omega_o) =
     \phi_i + \Frac{a}{\omega_i} ,
\eeq{e:e22}
  where index $i$ runs over frequencies and index $k$ runs over accumulation
periods.

  A solution of the $2\times 2$ system of normal equations that originates
from \Note{equations \ref{e:e22} can be easily obtained analytically.} Gathering
terms proportional to $a$, we express the contribution of the ionosphere
to phase and group delay as \Note{$\tau_p^{\rm iono} = - a / \omega_p^2$ and
$\tau_g^{\rm iono} =   a / \omega_g^2$}, where $\omega_p$ and $\omega_g$ are
effective ionospheric frequencies:
\beq
   \begin{array}{r@{\,}c@{\,}l}
   \omega_p &=& \sqrt { \omega_o \, \Frac
                 { \Sum_i^n w_i \cdot
                   \Sum_i^n w_i (\omega_i - \omega_o)^2  -
                   \lp \Sum_i^n w_i ( \omega_i - \omega_o ) \rp^2 \,
                 }
                 {
                   \Sum_i^n w_i (\omega_i - \omega_o)
                   \Sum_i^n w_i \frac{(\omega_i - \omega_o)}{\omega_i} -
                   \Sum_i^n w_i (\omega_i - \omega_o)^2 \cdot
                   \Sum_i^n \frac{w_i}{\omega_i}
                 } }
   \vhex\vhex\vhex \hspace{-2em} \\
   \omega_g &=& \sqrt \Frac
                 { \Sum_i^n w_i \cdot
                          \Sum_i^n w_i ( \omega_i - \omega_o)^2  -
                      \lp \Sum_i^n w_i ( \omega_i - \omega_o ) \rp^2 \,
                 }
                 {
                   \Sum_i^n w_i ( \omega_i - \omega_o )^2
                   \Sum_i^n \frac{w_i}{\omega_i} -
                   \Sum_i^n w_i \cdot
                   \Sum_i^n w_i \frac{(\omega_i - \omega_o)}{\omega_i} .
                 }
   \end{array},
\eeq{e:e23}
   \Note{where $w_i$ is the weight assigned to the fringe phase at the $i$th
frequency channel.}

  \Note{They have a clear physical meaning: if the wide-band signal was replaced
by a quasi-monochromatic signal with a group or phase effective ionosphere
frequency, then the contribution to group or phase delay of the wide-band
signal would be the same as the contribution of the quasi-monochromatic
signal at those effective frequencies.}

  For computing the contribution of the ionosphere, we used the total
electron contents (TEC) maps from analysis of linear combinations of
GPS observables made at two frequencies, 1.2276 and 1.57542~GHz. Using
GPS-derived TEC maps for data reduction of astronomic observations, first
suggested by \citet{r:ros00}, has became a traditional approach
for data processing. Analysis of continuous GPS observations from a global
network  comprising 100--300 stations makes it feasible to derive an empirical
model of the total electron contents over the span of observations using
the data assimilation technique. Such a model is routinely delivered by GPS
data analysis centers since 1998. For our analysis we used the TEC model
provided by the GPS analysis center CODE \citep{r:scha98}. The model gives
values of the TEC in zenith direction on a regular 3D grid with resolutions
$5\degr \times 5\degr \times 2^h$.

\begin{figure}[tbh]
   \centering
   \ifpre
       \includegraphics[width=0.5\textwidth,clip]{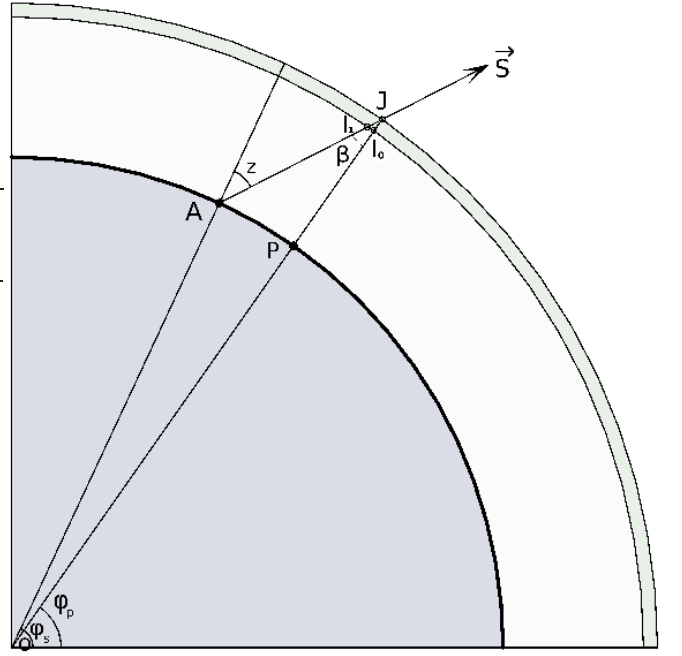}
   \else
       \includegraphics[width=0.5\textwidth,clip]{iono_1200.eps}
   \fi
   \caption{\note{Ray traveling from the source $\vec{s}$ to antenna $A$
            pierces the top of the ionosphere in the point $J$ and
            the bottom of the ionosphere in the point $I$. The ionosphere
            is considered as a thin layer. $P$ is the point on the Earth's
            surface beneath the ionosphere piercing point $J$.}
            }
   \label{f:iono}
\end{figure}

  For the purpose of modeling, the ionosphere is considered as a thin
spherical layer with constant height $H$ above the Earth's surface.
The typical value of $H$ is 450~km. In order to compute the TEC from GPS
maps, we need to know the coordinates of the point at which the ray pierces
the ionosphere --- point $J$ in figure~\ref{f:iono}. First, we find the
distance from the station to the ionosphere piercing point \Note{$D = |AJ|$
by solving triangle $OAJ$. Noticing that $|OA| = R_\earth$ and
$|OJ| = R_\earth +H$}, we immediately get
\beq
   \begin{array}{rcl}
     \beta & = & \arcsin \Frac{ \cos E }{1 + \frac{H}{R_\earth}} \vex \\
      D    & = & R_\earth \sqrt{
                  2 \Frac{H}{R_\earth} \lp 1 - \sin\lp E + \beta \rp\rp +
                    \lp\Frac{H}{R_\earth}\rp^2 } ,
   \end{array}
\eeq{e:e24}
  \Note{where $E$ is the elevation of the source above the horizon.}

  Then the Cartesian coordinates of point $J$ are $\vec{r} + D\vec{s}$.
Transforming them into polar coordinates, geocentric latitude and longitude,
we get arguments for interpolation in the 3D grid. We used the 3-dimensional
B-spline interpolation by expanding the TEC field into the tensor products
of basic splines of the 3rd degree. Interpolating the TEC model output, we
get the TEC through the vertical path $|JI_o|$. The slanted path $|JI_1|$
is $|JI_o|/\cos{\beta}$. Therefore, we need to multiply the vertical
TEC by $1/\cos{\beta(E)}$, which maps the vertical path delay through
the ionosphere into the slanted path delay. Here we neglect the ray path
bending in the ionosphere. We also neglect Earth's ellipticity, since the
Earth was considered spherical in the data assimilation procedure of
the TEC model.

  Combining equations, we get the final expression for the contribution
of the ionosphere to path delay:
\beq
  \tau_{iono} = \pm \Frac{a}{4\, \pi^2 \, f^2_{\rm eff} } \:
                \mbox{TEC} \: \Frac{1}{\cos{\beta(E)}} \quad,
\eeq{e:e25}
  where $f_{\rm eff}$ is the effective {\it cyclic} frequency, and the plus
sign is for the group delay, and the minus sign is
for the phase delay.

  Computation of the theoretical path delay and its partial derivatives over
model parameters is made with using software
VTD\footnote{Available at {\tt \url{http://astrogeo.org/vtd}}.}.

\subsection{Astrometric Analysis: Parameter Estimation}

   Astrometric analysis is made in several steps. First,
each individual 24~hour session    is processed independently.
The parameter estimation model includes estimation of 1)~clock functions
presented as a sum of a 2nd degree polynomial and a linear spline
over 60 minutes; 2)~residual zenith atmosphere path delay
for each station presented as a linear spline; 3)~coordinates
of all stations, except a reference station; and 4) coordinates of the target
sources. The goal of the coarse solution is to identify and suppress
outliers. The main reasons for outliers are a)~errors in determining the
global maximum of the fringe amplitude during fringe search and
b)~false detections. Both errors decrease with increasing the SNR.
Because of this, we initially run our solution by restricting to
SNR $\geq$ 6, and then restore good detections with SNR in the range
of $[5, 6]$.

   \Note{After identifying outliers and removing them from our solution,
we apply estimates of the parameters of the a~posteriori model to outliers,}
which allows us to predict the path
delay with accuracy better than 500~ps, except for sources that had fewer
than two detections. Then we re-run the fringe search for outliers and restrict
the search window to $\pm 1000$~ps with respect to predicted delay. We also
lower the SNR detection limit to 4.8, since the number of independent
samples in the restricted search window, and therefore, the probability
of false detection at a given SNR is less. This procedure allows to restore
\Note{from 50 to 80\% of observations marked as outliers in the previous step.}
The weights of observables were computed as
$ w = 1/\sqrt{\sigma_o^2 + r^2(b)} $, where $\sigma_o$ is the formal
uncertainty of group delay estimates and $r(b)$ is the baseline-dependent
reweighting parameter that was evaluated in a trial solution to make the
ratio of the weighted sum of the squares of residuals to its mathematical
expectation to be close to unity using the technique similar to that used
for fine fringe search.

  Finally, we run a global VLBI solution that uses all available
observations to date, 7.56 million, from April 1980 through
August 2010 in a single LSQ run. The estimated parameters are
\begin{itemize}
       \item [---]{\it global} (over the entire data set): coordinates of
                          4924 sources, including target objects in VGaPS
                          campaign, positions and velocities of all stations,
                          coefficients of B-spline expansion that model
                          non-linear motion of 17 stations, coefficients
                          of harmonic site position variations of 48
                          stations at four frequencies: annual, semi-annual,
                          diurnal, semi-diurnal, and axis offsets
                          for 67 stations.

       \item [---]{\it local}  (over each session):
                          tilts of the local symmetric axis of the atmosphere
                          (also known as ``atmospheric azimuthal gradients'')
                          for all stations and their rates, station-dependent
                          clock functions modeled by second order polynomials,
                          baseline-dependent clock offsets, and the Earth
                          orientation parameters.

       \item [---]{\it segmented} (over 20--60 minutes): coefficients of
                          linear splines that model atmospheric path delays
                          (20 minutes segment) and clock functions
                          (60 minutes segment) for each station. The estimates
                          of clock functions absorb uncalibrated instrumental
                          delays in the data acquisition system.
\end{itemize}

  The rate of change for the atmospheric path delays and clock functions between
adjacent segments was constrained to zero with weights reciprocal to
$ 1.1 \cdot 10^{-14} $ and \mbox{$2\cdot10^{-14}$}, respectively, in order
to stabilize solutions. We apply no-net rotation constraints on the positions of
212 sources marked as ``defining'' in the ICRF catalogue \citep{r:icrf98}
that requires the positions of these source in the new catalogue to have
no rotation with respect to the position in the ICRF catalogue to preserve
continuity with previous solutions.

  The global solution sets the orientation of the array with respect to
an ensemble of $\sim\! 5000$ extragalactic remote radio sources.
The orientation is defined by the series of Earth orientation parameters
and parameters of the empirical model of site position variations over
30 years evaluated together with source coordinates. Common sources observed
in VGaPS as atmosphere and amplitude calibrators provide a connection between
the new catalogue and the old catalogue of compact sources.

\subsection{Astrometric analysis: assessment of weights of observations}

\begin{figure*}[t]
   \centering
   \includegraphics[width=0.48\textwidth,clip]{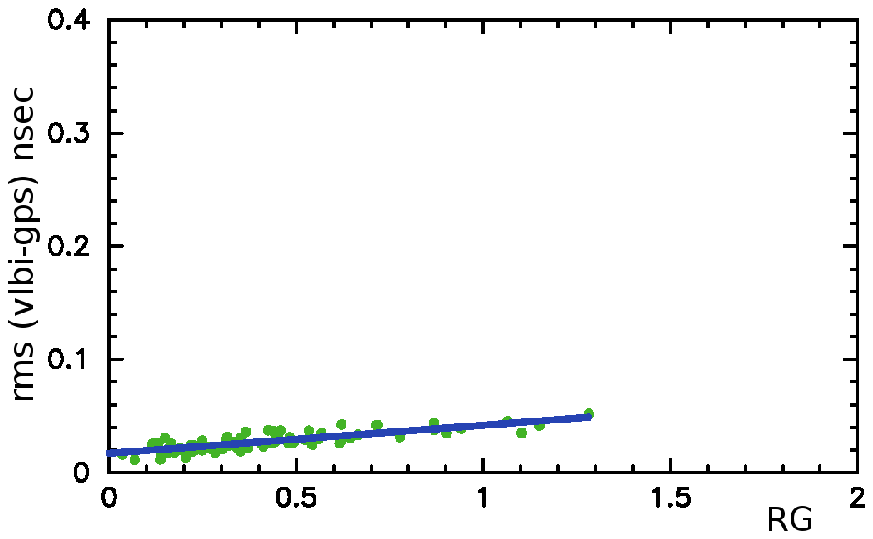}
   \hspace{0.02\textwidth}
   \includegraphics[width=0.48\textwidth,clip]{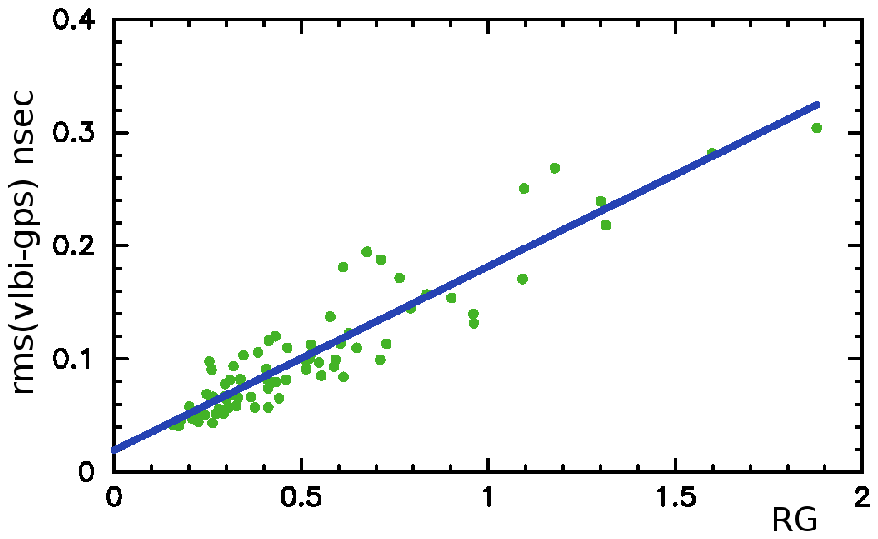}
\caption{Dependence of the rms of the differences VLBI minus GPS as
   a function of \note{ RG=$\sqrt{\mbox{rms}_{g1}^2 + \mbox{rms}_{g2}^2}$}
   for baseline {\sc fd-vlba/pietown}, $565$~km long (left)
   and for baseline {\sc la-vlba/mk-vlba}, $4\,970$~km long (right).
   The quantity $\mbox{rms}_{gi}$ is the rms of GPS path delay
   at the $i$ station of a baseline during a session. Each
   green dot corresponds to one VLBI session. The thick blue
   straight light is a linear fit through the data.
   \label{f:iono_rms}
}
\end{figure*}

\begin{figure}[htb]
   \centering
   \includegraphics[width=0.48\textwidth,clip]{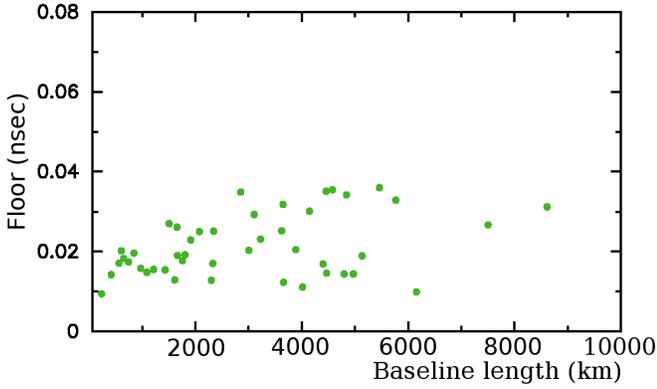}
   \caption{The floor of the regression model of the dependence of
            rms(VLBI--GPS) of $ RG = \sqrt{\mbox{rms}_{g1}^2 +
            \mbox{rms}_{g2}^2}$ for all VLBA baselines
            as a function of the baseline length.
            \label{f:iono_floor}
           }
\end{figure}

\begin{figure}[hbt]
   \centering
   \includegraphics[width=0.48\textwidth,clip]{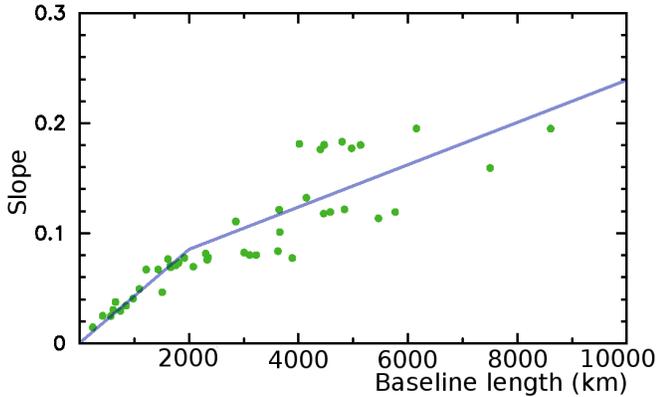}
   \caption{The slope of the regression model of dependence of
      rms(VLBI-GPS) of $ RG = \sqrt{\mbox{rms}_{g1}^2 +
      \mbox{rms}_{g2}^2}$ for all VLBA baselines
      as a function of the baseline length. \note{The straight lines
      show a linear approximation of the slope for two ranges
      of the baseline length: below and over 2000 km.}
      }
      \label{f:iono_slope}
\end{figure}

  As follows from the Gauss-Markov theorem, the estimate of parameters has
minimum dispersion when observation weights are chosen reciprocal to
the variance of errors. The group delays used in the analysis have errors
due to the thermal noise in fringe phases and due to mismodeling
theoretical path delay in the atmosphere:
\beq
      \sigma^2 = \sigma^2_{th} + \sigma^2_{io} + \sigma^2_{na}
\eeq{e:e26}
  where  $\sigma^2_{th}$ is thermal noise, $\sigma^2_{io}$ and $\sigma^2_{na}$
are the contribution of the ionosphere and the neutral atmosphere to the
error budget.

\subsubsection{A~priori errors of the GPS ionosphere model}
\label{s:iono_err}

  The first term, $\sigma^2_{th}$, was estimated during the fringe fitting.
The second term can only be evaluated indirectly. \citet{r:sek03} used
six dual-band intercontinental VLBI sessions at 10 stations in July 2000
and compared TEC values from GPS with TEC estimated from VLBI observables.
They drew a conclusion that the errors in path delay derived from the GPS
TEC model were at the range of 70~ps at the zenith direction at 8.6~GHz.
However, the ionosphere path delay is a non-stationary process. Therefore,
great caution should be taken in an attempt to generalize conclusions made
from analysis of a small network over a short time period.
The non-stationarity of ionospheric fluctuations implies that there
does not exist an exact expression for the variance of the ionosphere
fluctuations during any given period, and any expression for the variance
is an approximation.

  Since June 1998 when the GPS TEC maps became available through August 2010,
more than 2000 dual-band S/X VLBI sessions under geodesy and absolute
astrometry programs were carried out, including 92 sessions at the VLBA.
It can be easily shown that the contribution of the ionosphere at X-band,
$\tau_{xi}$ can be found from the linear combination of group delay
observables \Note{with coefficients that are expressed through effective
ionosphere frequencies at these bands $\omega_x$ and $\omega_s$
defined in expression \ref{e:e23}}:
\beq
      \tau_{xi} = (\tau_x - \tau_s) \,
                   \Frac{\omega^2_s}{\omega^2_x - \omega^2_s} .
\eeq{e:e27}

  We used this dataset for evaluation of the \Note{errors of the
contribution of the ionosphere to group delays derived from GPS TEC maps,
considering the ionosphere contribution from dual-band VLBI observations
as true for the purpose of this comparison. We computed the ionosphere contribution from the GPS model
and from VLBI observations for each session.} The root mean squares (rms)
of the differences of the contribution VLBI--GPS was computed for all
sessions and all baselines. We sought regressors that can predict
rms(VLBI--GPS). We expect the short-term variability of the ionosphere at
scales less than several hours to dominate the errors of the GPS model.
The sparseness of the GPS network and limited sky coverage result in missing
high frequency spatial and temporal variations of the ionosphere.
The turbulent nature of the ionosphere path delay variations suggests that
the rms of the ionosphere model errors due to missed high frequency
variations will be related to the rms of the low frequency variations either
as a linear function or as a power law. After many tries we found that
the following parameter can serve as a regressor:
RG = $\sqrt{\mbox{rms}_{g1}^2 + \mbox{rms}_{g2}^2}$, where $\mbox{rms}_{gi}$
is the rms of the GPS path delay at the $i$ th station of a baseline during
a session. The $\mbox{rms}_{g1}$ and $\mbox{rms}_{g2}$ are highly correlated
at short baselines and
$\sqrt{\mbox{rms}_{g1}^2 + \mbox{rms}_{g2}^2} > \mbox{rms(g2 - g1)}$.
At long baselines the ionosphere contribution de-correlates. Therefore, the
dependence of rms(VLBI--GPS) versus RG will depend on the baseline
length and possibly on other parameters. Figure~\ref{f:iono_rms} shows
examples of this dependence for a short baseline and for a long baseline.

  It is remarkable that rms(VLBI--GPS) versus RG fits reasonably well
with a linear function. We computed rms(VLBI--GPS) \Note{for each baseline
and fitted it to the linear model} $F + S \cdot \mbox{RG}$. The floor of
the linear fit has a mean value around 20~ps (Figure \ref{f:iono_floor}).
As expected, the slope of the fit increases with baseline length
as shown in Figure~\ref{f:iono_slope}. The growth is linear up to
baselines lengths of 2\,000~km, which apparently correspond
to the decorrelation of the paths through the ionosphere. The growth
of the slope beyond baseline lengths of 2\,000~km is slower
\Note{and it shows more scatter.}

  We use this dependence for predicting the rms of the GPS ionosphere
model errors. For each station that participated in the experiment we
computed the rms of the ionosphere variations in zenith direction.
Then we express the predicted variance of the GPS ionosphere model
errors as
\beq
   \begin{array}{lcl}
      \sigma_{i}^2 & = & \lp F_b \Frac{f^2_t}{f^2_{\rm eff}}\rp^2
             + \lp\tau_{iono,1} - \frac{\bar{\tau}^z_1}{\cos \beta(E_1)}
               \rp^2 S_b^2 \\
             & & + \lp\tau_{iono,2} - \frac{\bar{\tau}^z_2}{\cos \beta(E_2)}
               \rp^2 S_b^2
          , \hphantom{aaa}
    \end{array}
\eeq{e:e28}
  where $\tau_{iono,i}$ is the ionosphere path delay at the $i$ th station
computed using the GPS TEC maps, $\bar{\tau}^z_1$ the zenith ionosphere path
delay from GPS TEC maps averaged over a $24^h$ period with respect to the
central date of the session, $F_b$ and $S_b$ are parameters of the linear
model rms(VLBI--GPS) versus RG \Note{for a given baseline}, $f_t$ is the
frequency for which the model was computed (8.6~GHz),
\Note{and $f_{\rm eff}$ the frequency of the experiment for which the
model is applied (24.5~GHz).} The term
$\tau_{iono,i} - \frac{\bar{\tau}^z_i}{\cos \beta(E_i)}$
is the difference between the ionosphere path delay from GPS at a given
direction and the average ionosphere path delay scaled to take into account
the elevation dependence. \Note{This term is an approximation}
of ${\mbox{rms}_{g1}}$ used for computation of RG.


\subsubsection{A~priori errors of the path delay in the neutral atmosphere}

Rigorous analysis of the errors of modeling the path delay in the neutral
atmosphere is beyond the scope of this paper. Assuming the dominant
errors of the a~priori model are due to high frequency fluctuations
of water vapor at scales less than 3--5 hours, we seek the regression
model in the form of dependence of the rms of errors on the total path
delay in the non-hydrostatic component of the path delay. We made several
trial runs using all 123 observing sessions at the VLBA under geodesy and
absolute astrometry programs \Note{with reciprocal weights modified
according to}
\beq
   \sigma^2_{used} = \sigma^2 + \lp a \cdot \Frac{\tau_{s}}{\tau_z} \rp^2 .
\eeq{e:e29}

  Here ${\tau_s}$ is the contribution of the non-hydrostatic constituent
of the slanted path delay and ${\tau_z}$ is the non-hydrostatic path
delay in the zenith direction computed by direct integration of the equations
of wave propagation through the atmosphere using the refractivity computed
from  the MERRA model, and $a$ is the coefficient. We found that
when coefficient $a=0.02$ is used, the baseline length repeatability,
defined as the rms of the deviation of baseline length with respect to
the linear time evolution, reaches the minimum. We adopted value $0.02$
in our analysis of VGaPS experiments. For typical values of
${\tau_z}$, the added noise is 8~ps in zenith direction and 80~ps at
$10^\circ$ elevation.

\subsubsection{Ad hoc added variance of the noise}

  We also computed for each baseline and each session an ad hoc variance
of observables that, added in quadrature, makes the ratio of the weighted
sum of squares of post-fit residuals to their mathematical expectation close
to unity in a similar way as we updated fringe phase weights.
Expression~\ref{e:e14} was used for computing this variance. This ad hoc
variance was applied to further inflate the a~priori observable
uncertainties that have already been corrected for the inaccuracy of the
a~priori model of wave propagation through the ionosphere and the neutral
atmosphere according to expressions \ref{e:e28} and \ref{e:e29}. In contrast
to \ref{e:e28} and \ref{e:e29}, the baseline-dependent ad hoc variance
is elevation independent.

\section{Validation of the wide-band fringe search algorithm}
\label{s:validation}

  Using \PIMA, we detected 327 target sources versus 136 targets
detected with AIPS, since the AIPS detection limit is lowered by a
factor of $\sqrt{8} = 2.83$.  Thus, the yield of the experiments was
improved by the factor of 2.4 -- a very significant improvement that
well justified our efforts to create a new software package for
processing astrometric observations.

\begin{figure}[t!]
  \centering
  \ifpre
      \includegraphics[width=0.48\textwidth,clip]{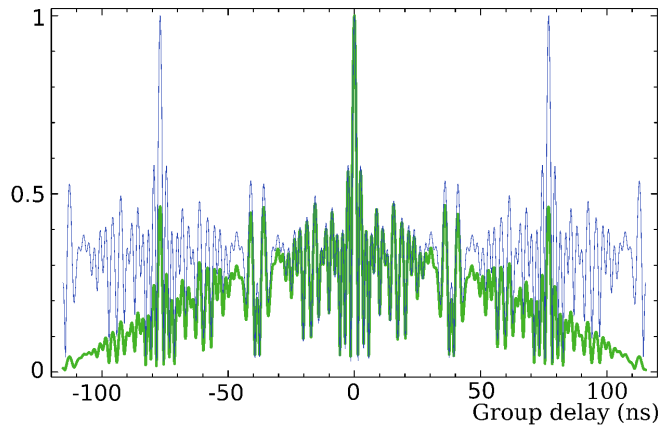}
  \else
      \includegraphics[width=0.48\textwidth,clip]{drf_1200.eps}
  \fi
  \caption{Coherent sum of cross-correlation amplitude, normalized to
           its value at the maximum as a function of group delay shift
           with respect to the maximum for VGaPS observing sessions.
           The thick green line shows result of processing with \PIMA.
           The thin blue line shows results of processing with AIPS.
           The pattern of the thin blue line repeats with a period of 76.923~ns.
  \label{f:drf}
  }
\end{figure}

Another difference is that the results of processing the data with
AIPS have ambiguities in group delay that are reciprocal to the
minimum difference between intermediate frequencies. This means that
the group delays are $\tau_g + K \tau_s$, where $\tau_s = 76.923$~ns
and $K$ is {\it an arbitrary} integer number. At the same time, the results
of processing with the wide-band fringe fitting algorithm do not suffer
this problem. The reason for group delay ambiguities is that the
narrow-band fringe fitting algorithm implemented in AIPS first
coherently averages the data within each IF, and in the second step of
fringe fitting it processes a rail of narrow-band signals. The Fourier
transform that describes the dependency of the amplitude of the coherent
sum of the cross-spectrum on group delay has a periodicity that is
reciprocal to the minimum frequency separation of IFs, 76.923~ns in our case.
The wide-band fringe fitting algorithm does not average the
spectrum. Therefore, the periodicity of the Fourier transform of the
coherent sum of the cross-spectrum in the wide-band algorithm is equal
to the spectral resolution, i.e. 2~mks for VGaPS experiments.
Figure~\ref{f:drf} illustrates this difference.

  The lack of group delay ambiguities has a profound effect on determining
source positions with poorly known a~priori coordinates. In the presence
of group delay ambiguities, we had to solve for source positions first using
less precise so-called narrow-band group delays determined by arithmetic
averaging group delays computed for each IF independently. The accuracy of
these source position estimates is often insufficient to reliably
resolve group delay ambiguities, especially in the presence of narrow-band
group delay outliers. When the number of used observations is large, say more
than 10, the data redundancy allows us to detect the presence of incorrectly
resolved ambiguities and fix the problem. But if the number of observations
is small, chances are the error in group delay ambiguity resolution will
not be noticed. In the past, source position estimates made with less than
8 observations were considered unreliable. The use of the wide-band fringe
fitting algorithm eliminates this problem entirely. Reprocessing the old
data revealed that group delay ambiguities for a considerable number
of sources were indeed resolved incorrectly, which resulted in source
position errors as large as $4'$! In contrast, the wide-band fringe fitting
algorithm provides reliable estimates of source positions using
a minimum redundancy of 3 observations.

   Since the wide-band fringe search algorithm is new, we would like
to be sure that the new algorithm does not introduce new systematic
errors with respect to the old one. As a validation test, we re-processed
a set of 33 VLBA absolute astrometry/geodesy experiments under the
RDV program \citep{r:rdv} and 12 VLBA absolute astrometry experiments under
the K/Q program \citep{r:kq}. Each test experiment  had a duration
of 24 hours.

   The RDV experiments were observed on a global network, including all 10
VLBA stations, with dual-band receivers at 8.6~GHz (X~band) and
2.3~GHz (S~band), with 4 IFs allocated at X band and 4 IFs allocated at
S band. Fringe fitting, outlier elimination, re-weighting, and in the
case of AIPS, group delay ambiguity resolution, were made independently
using \PIMA and AIPS. Subsequent data reduction and parameter estimation
were made using identical setups. Estimated parameters of the solution
were the same as in processing the VGaPS sessions, except for treatment
of site positions: they were treated as local parameters, i.e.
estimated for each session independently.

\begin{table}[t!]
  \caption{\note{Solution statistics from 33 global RDV sessions
           processed with AIPS and \PIMA. The statistics in
           the central column were computed using all observations.
           The statistics in the right column were computed using
           observations with \mbox{SNR $> 10$.}}}
  \centering
  \begin{tabular}{l @{\qquad} l @{\qquad} l @{\quad} l}
     \hline\hline
               &  AIPS & \PIMA  & \PIMA  \\
               &    & SNR${}_{\rm min} = 5.0$ &
                      SNR${}_{\rm min} = 10.0$  \hspace{-1em} \vhex \\
     \hline
     No.\ obs used &  467\,769     & 531\,299            & 472\,717  \\
     fit wrms    &  18.40 ps     & 21.22 ps            & 19.65 ps \\
     No.\ sources  &  776          & 800                 & 773       \\
     wrms $ \Delta \Psi \cos \epsilon_o $ & 0.10 mas & 0.12 mas & 0.14 mas \\
     wrms $ \Delta \epsilon $             & 0.10 mas & 0.10 mas & 0.12 mas \\
     Bas. rep. at 5000 km & 4.81~mm & 4.75~mm          & 4.96~mm   \\
     Bas. rep. at 9000 km & 8.54~mm & 8.08~mm          & 7.95~mm   \vhex \\
     \hline
  \end{tabular}
  \label{t:rdv_diff}
\end{table}

The statistics of the solution for 33 global RDV sessions using AIPS
and \PIMA are shown in table~\ref{t:rdv_diff}. The weighted root mean
squares (wrms) of the postfit residuals is larger in the \PIMA solution
for two reasons. First, the \PIMA solution contain 14\% more points, mainly
with SNR's in the range [5.0, 10.0], that were undetected by the
traditional AIPS algorithm. \Note{The errors of these observables
are systematically larger.} We rerun the \PIMA solution
and excluded all points with SNR either at X- or S-band less than
10. The difference in wrms postfit residuals was significantly
reduced. The second reason is that the group delay formal errors were
inflated in \PIMA processing to make the ratio of postfit residuals of
fringe fitting to its mathematical expectation close to 1.  This was
not done for the AIPS solutions.

\begin{figure*}[thb]
  \centering
  \ifpre
     \includegraphics[width=0.48\textwidth,clip]{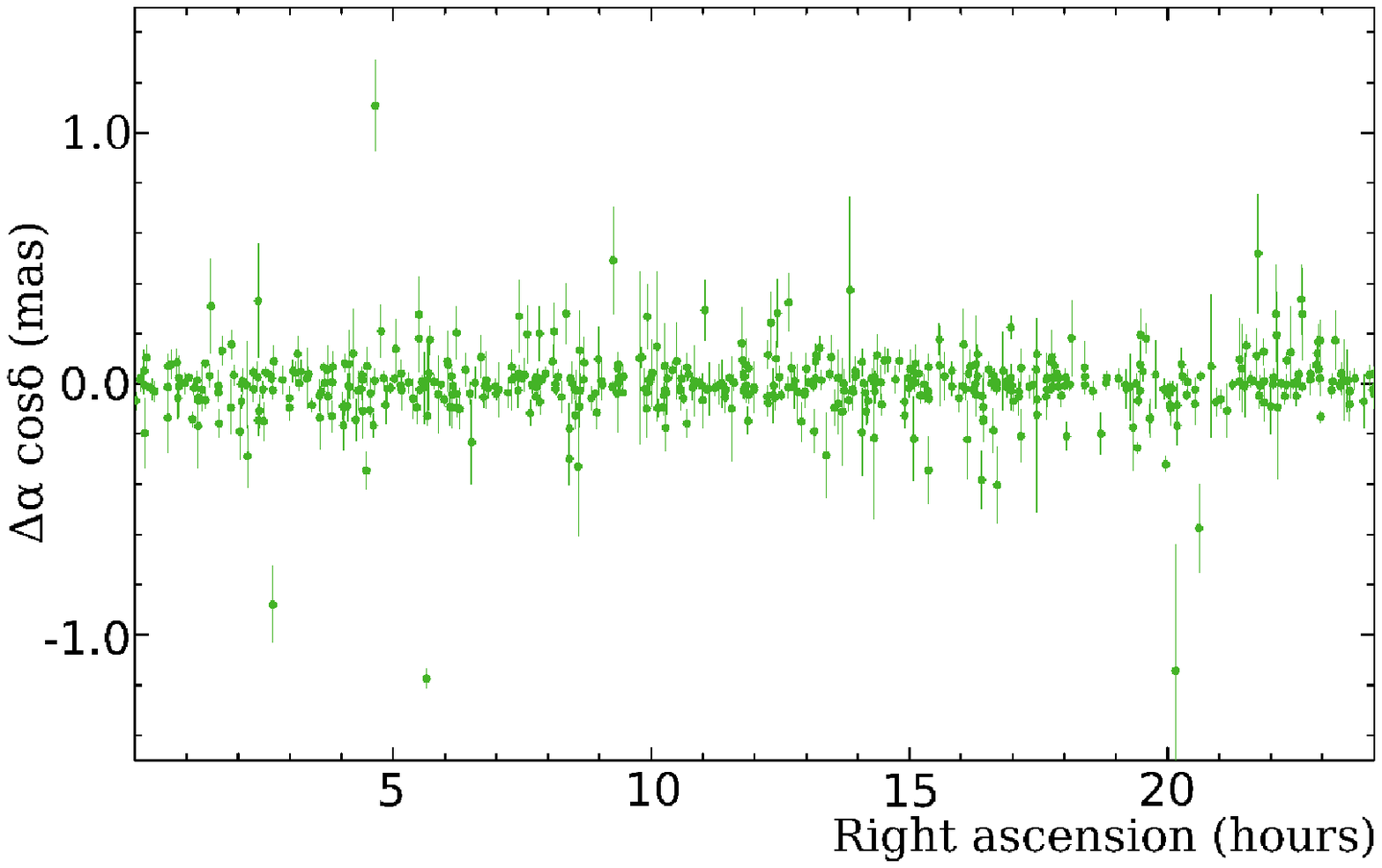}
  \else
     \includegraphics[width=0.48\textwidth,clip]{delta_alpha_1200.eps}
  \fi
  \hspace{0.02\textwidth}
  \ifpre
      \includegraphics[width=0.48\textwidth,clip]{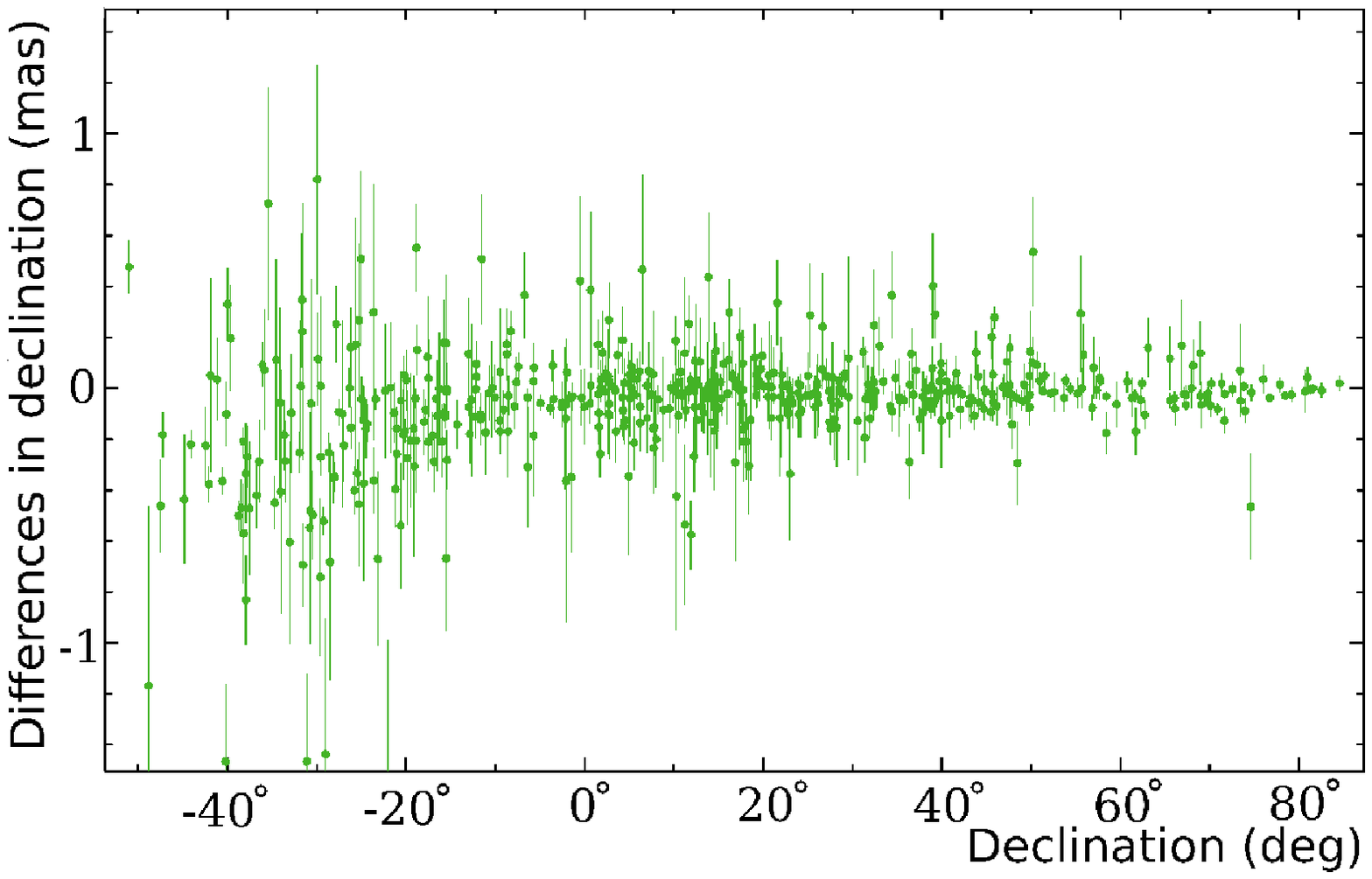}
  \else
      \includegraphics[width=0.48\textwidth,clip]{delta_delta_1200.eps}
  \fi
  \caption{Systematic differences in source coordinate estimates between
           solutions using group delays derived by \PIMA and by
           AIPS. Left plot shows $\Delta \alpha \; \cos \delta \; (\alpha)$,
           right plots shows $\Delta \delta(\delta)$.
  \label{f:syserr1}
  }
\end{figure*}

Another test of goodness of the solution and possible presence of
systematic errors is the so-called baseline length repeatability test
\citep{r:rdv}.  We computed the wrms of baseline length estimates with
respect to the fitted linear model of their evolution with time. The
dependence of the baseline length wrms with the length of baselines
$L$ is fitted by function $R(L) = \sqrt{A^2 + B^2 \, L}$. Values of
R(L) at L=5000~km and L=9000~km are presented in
table~\ref{t:rdv_diff}. We also computed the wrms of the deviations of
estimates of daily offsets of nutation angles in longitude,
$\Delta\psi$, and nutation in obliquity, $\Delta\epsilon$, with
respect to the empirical nutation expansion {\tt heo\_20091201}.  The
statistics in table~\ref{t:rdv_diff} show the satisfactory agreement
between AIPS and \PIMA solutions and do not reveal any systematic
differences.

  Since the goal of VGaPS was to derive source positions, comparison of
the positions from AIPS and \PIMA processing is important \Note{to
evaluate the level of systematic differences}.  We, thus, computed the
differences in sources coordinates $\Delta\alpha \cos \delta ( \alpha
)$ and $\Delta\delta(\delta)$.  We restricted our analysis to 528
sources that had more than 64 observations in both AIPS and \PIMA in
order to avoid effects of a greater number of observations available
in the \PIMA solutions.  Plots of these differences are shown in
figure~\ref{f:syserr1} and position comparisons are shown in
table~\ref{t:rdv_sou_diff}.  The table contains the results from 528
sources observed in RDV sessions \Note{with at least 64 used
observations.} The 2nd column shows statistics of differences between
AIPS and \PIMA with an SNR cutoff of 5. The 3rd column shows the
differences between AIPS and \PIMA solutions with an SNR cutoff of 10.
\Note{These differences are only about 20\% of a typical position
uncertainty.}

\begin{table}[t!]
  \caption{\note{Differences between AIPS and \PIMA positions of 528
           sources observed in RDV experiments. The differences in
           the left column were computed using all observations.
           The differences in the right column were computed using
           observations with \mbox{SNR $> 10$.}}}

  \centering
  \begin{tabular}{l @{\qquad} c @{\qquad} c}
     \hline\hline
               &  AIPS--\PIMA  &
                  AIPS--\PIMA  \vhex \\
               &  SNR${}_{\rm min} = 5.0$ &
                  SNR${}_{\rm min} = 10.0$ \hspace{-1em}  \vhex \\
     \hline
     wrms $\Delta\alpha \cos \delta (\alpha)$ &
          0.072 mas & 0.020 mas               \\

     wrms $\Delta\delta(\delta)             $ &
          0.088 mas & 0.032 mas               \\

     \hline
  \end{tabular}
  \label{t:rdv_sou_diff}
\end{table}

\begin{figure}[thb]
  \centering
  \ifpre
      \includegraphics[width=0.48\textwidth,clip]{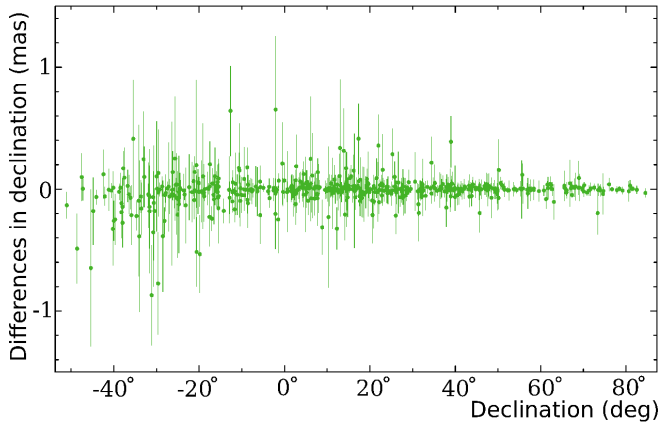}
  \else
      \includegraphics[width=0.48\textwidth,clip]{delta_delta_snr_12_1200.eps}
  \fi
  \caption{Systematic differences $\Delta \delta(\delta)$ in source
           coordinate estimates between solutions using group delays
           derived by \PIMA with SNR cutoff 10 and by AIPS.
  \label{f:syserr2}
  }
\end{figure}

In the right plot of Figure \ref{f:syserr1} we can see a small
systematic declination differences at low declinations.
\Note{Similar differences but produced from \PIMA solution made with the
SNR cutoff 10.0, shown in Figure~\ref{f:syserr2}, help to understand
the origin of this pattern. When the SNR cutoff is raised to 10, the
wrms of differences are reduced by 2--3 times to 0.02--0.03~mas
and the systematic pattern disappear. For comparison, the average formal
uncertainties for declinations are 0.12~mas and for right ascensions
scaled by $\cos\delta$ are 0.07~mas, so the systematic errors are small.
Sources at low declinations observed on a VLBI array located in the northern
hemisphere are necessarily taken at low elevations. Including in a solution
additional low SNR observations at low elevations, unavailable in the AIPS
solutions, changes the contribution of systematic errors due to mismodeling
path delay in the neutral atmosphere that are higher at low elevations.
At present, it is not clear whether including observations with SNR in range
5--10 increases systematic errors, or the opposite, including these
observations reduces the systematic errors. However,  the magnitude of the
differences, less than 0.5~mas as declinations in the range
$[-50\degr\!, -25\degr]$ does not raise a concern.}

Finally, we have re-analyzed 12 VLBA experiments under the K/Q program.
The frequency setup in K/Q and VGaPS campaigns was identical. Lowering
the detection limit by a factor of $\sqrt{8} \approx 2.83$ with the
use of the wide-field fringe fitting algorithm greatly
helped. Processing the data with \PIMA\ did not detect only 8 out of
the 340 observed sources, with 61 non-detections in data processing
with AIPS. We compiled table~\ref{t:kq_diff} of statistics of the K/Q
solution in a form similar to table~\ref{t:rdv_diff}. Analysis of
source position differences did not reveal any pattern of systematic
errors. The statistics of these differences presented in
table~\ref{t:kq_sou_diff} do not exceed formal uncertainties of
positions which are 0.08~mas and 0.14~mas in right ascensions scaled
by $\cos\delta$ and declination, respectively.  The first column is
four sources with an SNR cutoff of 5, the second column with an SNR
cutoff \Note{of $5 \cdot \sqrt{8} = 14.1$. The baseline length
repeatability and the wrms of nutation offset time series from \PIMA
solutions are 20--30\% smaller. We do not have an explanation why
\PIMA solution produces noticeably better results for 24~GHz
observations, but no significant improvement was found from analysis
of S/X RDV experiments.}

\begin{table}[b!]
  \caption{\note{Solution statistics from 12 VLBA experiments at
           24~GHz under the K/Q program processed with AIPS and \PIMA.
           The statistics in the central column were computed using
           all observations. The statistics in the right column were
           computed using observations with \mbox{SNR $> 10$.}}}
  \centering
  \begin{tabular}{l @{\qquad} l @{\qquad} l @{\quad} l}
     \hline\hline
               &  AIPS & \PIMA  & \PIMA  \\
               &    & SNR${}_{\rm min} = 5.0$ &
                      SNR${}_{\rm min} = 14.1$ \hspace{-1em}  \vhex \\
     \hline
     \# Obs used &  104\,887  & 139\,213  & 105\,761  \\
     fit wrms    &  19.50 ps  & 23.08 ps  & 19.19 ps  \\
     \# Sources  &  279       & 332       & 276       \\
     wrms $ \Delta \Psi \cos \epsilon_o $ & 0.13 mas & 0.10 mas & 0.09 mas \\
     wrms $ \Delta \epsilon $             & 0.18 mas & 0.13 mas & 0.14 mas \\
     Bas. rep. at 5000 km & 5.56~mm & 4.58~mm          & 4.58~mm   \\
     Bas. rep. at 9000 km & 9.29~mm & 7.36~mm          & 7.32~mm   \vhex \\
     \hline
  \end{tabular}
  \label{t:kq_diff}
\end{table}

\begin{table}[b!]
  \caption{\note{Differences between AIPS and \PIMA positions of 244
           sources observed in the K/Q VLBA experiments. The differences in
           the left column were computed using all observations.
           The differences in the right column were computed using
           observations with \mbox{SNR $> 10$.}}}
  \centering
  \begin{tabular}{l @{\qquad} c @{\qquad} c}
     \hline\hline
               &  AIPS--\PIMA  &
                  AIPS--\PIMA  \vhex \\
               &  SNR${}_{\rm min} = 5.0$ &
                  SNR${}_{\rm min} = 10.0$ \hspace{-1em}  \vhex \\
     \hline
     wrms $\Delta\alpha \cos \delta (\alpha)$ &
          0.062 mas & 0.060 mas               \\

     wrms $\Delta\delta(\delta)             $ &
          0.108 mas & 0.096 mas               \\

     \hline
  \end{tabular}
  \label{t:kq_sou_diff}
  \par\vspace{-0.5ex}
\end{table}

The above results of analysis of validation runs processing 0.6
million observations at K, X and S bands, collected during 1080 hours
of recording at the VLBA and the global VLBI network with both AIPS
and \PIMA demonstrate that the new wide-band algorithm for group
delays does not introduce any significant systematic errors while
detecting more sources because \Note{it evaluates group delays using
the coherent sum of the data across the wide-band}. We conclude that
\PIMA has passed \Note{major validation tests.}


\section{Investigation of systematic errors in source positions}
\label{s:errors}

Using single-band data, ionosphere path delay mismodeling may produce
systematic position errors. \citet{r:kq} showed that in their analysis
of K-band observations systematic errors reached several mas and had a
tendency to be larger at low declinations.  In
section~\ref{s:iono_err} we evaluated the rms of random errors caused
by ionosphere path delay mismodeling. However, inflating weights to
account for the variance of errors in general does not guarantee that
\Note{source positions will have no systematic errors.}

\begin{figure}[thb]
   \centering
   \includegraphics[width=0.48\textwidth,clip]{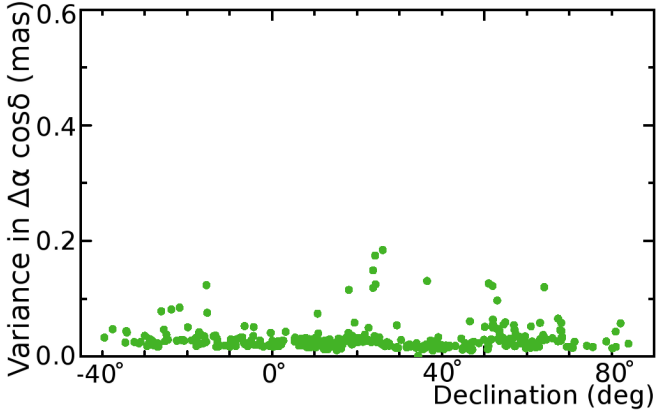}
   \caption{Modeled systematic errors $\Delta \alpha_i(\delta) \cos \delta$
            driven by the mismodeling ionosphere path delay
            contribution evaluated from the Monte Carlo simulation.
           \label{f:iono_alpha_error}
           }
\end{figure}

\begin{figure}[ht]
   \centering
   \includegraphics[width=0.48\textwidth,clip]{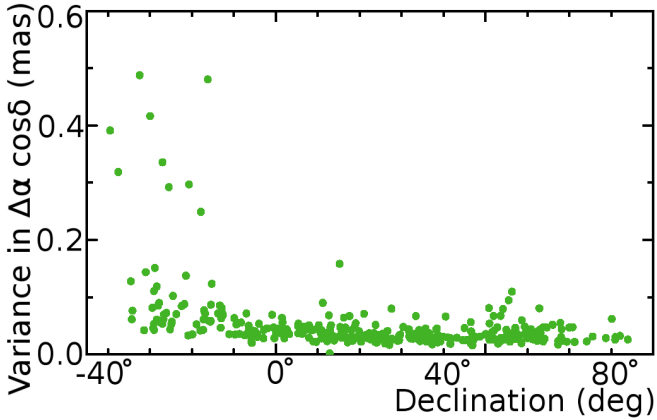}
   \caption{Modeled systematic errors $\Delta \delta^i(\delta)$ driven by
            the mismodeling ionosphere path delay contribution evaluated
            from the Monte Carlo simulation.
           \label{f:iono_delta_error}
           }
\end{figure}

  To evaluate the magnitude of possible ionosphere driven systematic
errors we made the following Monte Carlo simulation. We added to
the theoretical path delay the zero-mean Gaussian noise
$M \cdot N(0,\sigma_i)$, with the variance $\sigma_i$ computed according
to expression~\ref{e:e28}. The noise was magnified  $M$ times in order
to make the contribution of ionosphere path delay errors dominant over other
sources of errors. We made 64 analysis runs of VGaPS data using different
seeds of the random noise generator. The magnification factor 100 was used.
Thus, we produced 64 estimates of position of each source with added noise.
We computed the rms of position estimates of each target source
and divided it by $M$. To the extent of the validity
of expression~\ref{e:e28}, these rms's represent expected errors due to
the inadequacy of the ionosphere path delay models based on using the
TEC models derived
from GPS analysis. Plots of $\Delta \alpha_i(\delta)$ and
$\Delta \delta_i(\delta)$ errors are shown in
Figures~\ref{f:iono_alpha_error}--\ref{f:iono_delta_error}. For 90\%
of the sources, errors are at the level of 0.02--0.04~mas.
$\Delta \delta(\delta)$ increases to 0.1~mas at declinations less than
-$20^\circ$, and for some sources may reach 0.4~mas. The disparity
in systematic errors for sources at comparable declinations reflects the
disparity in the number of observables used in the solution.

  For assessment of remaining systematic errors we exploited that fact that
56 known sources were observed as amplitude and atmospheric calibrators.
Positions of these sources are known from previous dual-band S/X observations
with accuracies better than 0.1~mas. We split the set of 56 calibrators into
two subsets of 28 objects and ran two additional solutions. In the first
solution we suppressed 28 calibrators in all sessions except VGaPS and
determined their positions solely from VGaPS. In the second
solution we did the same with the second subset. Considering that the
positions of calibrators from numerous S/X observations can be regarded
as true for the purpose of this comparison, we treated the differences as VGaPS errors.

  We computed the \Note{$\chi^2$ per degree of freedom statistics for}
the differences in right ascensions and declinations $\Delta \alpha$ and
$\Delta \delta$ and sought additional variances $v_\alpha$ and $v_\delta$
that, being added in quadrature to the source position uncertainties,
will make them close to unity:
\beq
   \begin{array}{l @{\enskip} c @{\enskip} l}
      \frac{\chi^2_\alpha}{\mbox{ndf}} & = &
            \Frac{\sum_{k=1}^{k=n} \Delta \alpha^2_k \cos^2\delta_k}
            {n \, \sum_{k=1}^{k=n} \sqrt{ \sigma^2_{\alpha,k} \cos^2 \delta_k +
                   \alpha_{i,k}^2 + v^2_\alpha \cos^2 \delta_k} }
      \vex \\
      \frac{\chi^2_\delta}{\mbox{ndf}} & = &
            \Frac{\sum_{k=1}^{k=n} \Delta \delta^2_k }
           {n \,  \sum_{k=1}^{k=n} \sqrt{ \sigma^2_{\delta,k} +
                                          \delta_{i,k}^2  + v^2_\delta} }
   \end{array}
\eeq{e:e30}

  \Note{The denominator in equation \ref{e:e30} is a mathematical
expectation of the sum of squares of differences, provided the estimates
of source positions are statistically independent.}

  We found the following additive corrections of the uncertainties
in right ascensions scaled by $\cos \delta$ and for declinations
respectively: $v_\alpha = 0.08$~mas and $v_\delta$=0.120~mas.

  The final inflated errors of source positions, $\sigma^2_\alpha(f)$ and $
\sigma^2_\delta(f)$, are
\beq
   \begin{array}{l @{\enskip} c @{\enskip} l}
      \sigma^2_\alpha(f) = \sigma^2_\alpha + v^2_\alpha +
                           \alpha^2_i/\cos^2 \delta
      \vex \\
      \sigma^2_\delta(f) = \sigma^2_\delta + v^2_\delta + \delta^2_i
   \end{array}
\eeq{e:e31}

\begin{figure*}[t]
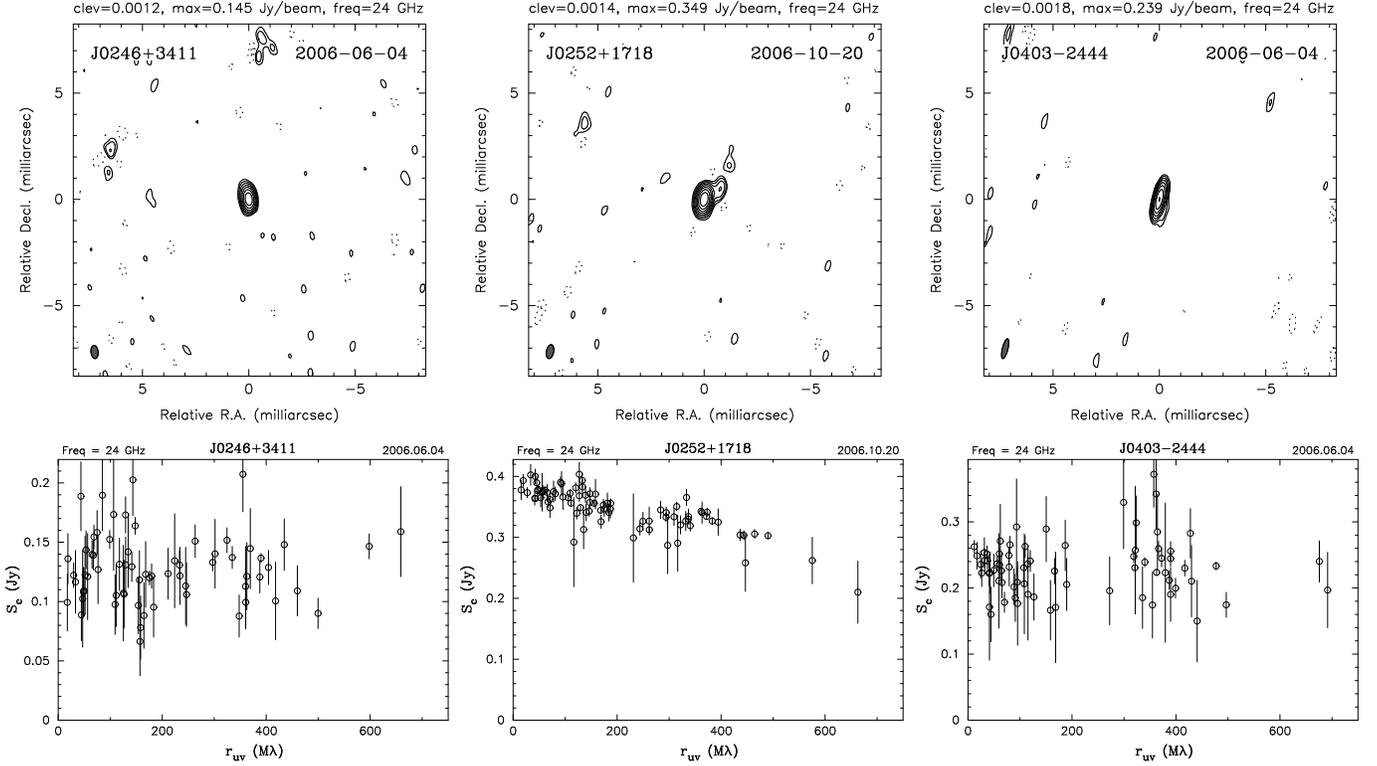

\centering
\resizebox{1.0\hsize}{!}
{
  \includegraphics[trim=-0.5cm 6cm 0cm 0cm]{J0246+3411_K_2006_06_04_yyk_map.ps}
  \includegraphics[trim=-1.0cm 6cm 0cm 0cm]{J0252+1718_K_2006_10_20_yyk_map.ps}
  \includegraphics[trim=-1.0cm 6cm 0cm 0cm]{J0403-2444_K_2006_06_04_yyk_map.ps}
}

\resizebox{1.0\hsize}{!}
{
  \includegraphics[trim=-1cm  0.0cm 0cm 0cm,angle=270]{J0246+3411_K_2006_06_04_yyk_rad.ps}
  \includegraphics[trim=-1cm -0.5cm 0cm 0cm,angle=270]{J0252+1718_K_2006_10_20_yyk_rad.ps}
  \includegraphics[trim=-1cm -0.5cm 0cm 0cm,angle=270]{J0403-2444_K_2006_06_04_yyk_rad.ps}
}
\caption{
From top to bottom.
{\em Row~1:}
Naturally weighted CLEAN images at 24~GHz. The lowest contour
is plotted at the level given by ``clev'' in each panel title (Jy/beam),
the peak brightness is given by ``max'' (Jy/beam). The contour levels
increase by factors of two. The dashed contours indicate negative flux.
The beam is shown in the bottom left corner of the images.
{\em Row~2:}
Dependence of the correlated flux density on projected
spacing. Each point represents a coherent
average over one 2~min observation on an individual interferometer
baseline. The error bars represent only the statistical errors.
\label{f:images}
}
\end{figure*}

Position of some sources may indeed be different due to the core-shift
effect \citep{r:kov08,r:por09}. Treatment of this shift as VGaPS errors
makes our estimates of reweighting parameters and therefore reported final
inflated errors somewhat too conservative.

\section{Imaging}
\label{s:images}


  For imaging purposes we performed the a~priori calibration of
the data following a traditional way, using AIPS \citep{r:aips}. In the future,
we plan to introduce all extra steps required for an accurate amplitude
calibration into \PIMA as well.

  We followed the usual AIPS initial VLBA calibration procedure
involving a priori amplitude calibration with measured antenna gain
curves and system temperatures as well as sampling-based calibration
adjustments. Atmospheric absorption is significant at 24~GHz. We have
estimated its effect using system temperature data covering the whole
range of elevations and weather information in order to adjust
visibility amplitudes for the opacity. Typical values of the opacity
were found to range between 0.03 and 0.1 for different VLBA
telescopes and observing epochs. We performed phase calibration using
the phase calibration signal injected during observations and fringe
fitting. A separate solution for \YNote{station-based group delay and
phase delay rate was made for each frequency channel (IF)}. As the
final step of calibration, bandpass corrections were determined
and applied.

Our observations were scheduled around 24~GHz since the K-band
continuum performance is better at this frequency, away from the water
line.  Unfortunately, at the time of these observations most of the
VLBA telescopes had no gain curve measurements close to 24~GHz. It has
changed since 2007 when regular 24~GHz gain curve measurements started
to be performed at all VLBA stations. For all VLBA antennas but {\sc
  mk-vlba} and {\sc hn-vlba} we have used gain curves measured at
22.2~GHz while for the former the curves at 23.8~GHz were
applied. Antenna efficiency as well as the noise diode spectrum do
change with frequency \citep[see, e.g.,][as well as results of VLBA
gain curve measurements at 22 and 24 GHz after
2007\footnote{\tt \url{http://www.vlba.nrao.edu/astro/VOBS/astronomy/
\ifpre \else\relax\fi%
vlba\_gains.key}}]{r:vcs6}. This is one of the main sources of the
total amplitude calibration uncertainty.
Additionally, IFs in our experiment are widely spread
(Table~\ref{t:frq}) which might introduce extra amplitude shifts. We
used strong flat-spectrum sources in the sample in order to
estimate global relative amplitude correction factors for different IFs
but did not find any to be larger than 10\% with high confidence. No
extra frequency channel specific amplitude corrections were applied to
the data.

After a~priori calibration, data were imported to the Caltech DIFMAP
package \citep{r:difmap}, visibility data flagged, and images were
produced using an automated hybrid imaging procedure originally
suggested by Greg Taylor \citep{r:difmap-script} which we optimized
for our experiment. The procedure performs iterations of phase and
amplitude self-calibration followed by CLEAN image reconstruction. We
were able to reach the VLBA image thermal noise level for most of our
final CLEAN images. Examples of 3 images, for compact and resolved
objects, are shown in Figure~\ref{f:images}.

\ifpre
  \begin{deluxetable*}{ c l l l r r r r r r r r r l l}
\else
  \begin{deluxetable}{ c l l l r r r r r r r r r l l}
\fi
\tablecaption{The VGaPS catalogue \label{t:cat}}
\tabletypesize{\scriptsize}
\centering

\ifpre \relax \else \rotate \fi
\tablehead{
\vspace{2.0ex} \\
   \colhead{} &
   \multicolumn{2}{c}{Source name}                   &
   \multicolumn{2}{c}{J2000.0 Coordinates}           &
   \multicolumn{3}{c}{Errors (mas)}                  &
   \colhead{}                                        &
   \multicolumn{2}{c}{Correlated flux }              &
   \colhead{}
   \vspace{0.5ex} \\
   \colhead{} &
   \colhead{} &
   \colhead{} &
   \colhead{} &
   \colhead{} &
   \colhead{} &
   \colhead{} &
   \colhead{} &
   \colhead{} &
   \multicolumn{2}{c}{density (in Jy)}              &
   \vspace{0.5ex} \\
   \multicolumn{2}{c}{}         &
   \vspace{0.5ex} \\
   \colhead{Class}    &
   \colhead{IVS}      &
   \colhead{IAU}      &
   \colhead{Right $\,$ ascension} &
   \colhead{Declination}     &
   \colhead{$\Delta \alpha$} &
   \colhead{$\Delta \delta$} &
   \colhead{Corr}   &
   \colhead{\# Obs} &
   \colhead{Total } &
   \colhead{Unres }
   \vspace{0.5ex} \\
   \colhead{(1)}    &
   \colhead{(2)}    &
   \colhead{(3)}    &
   \colhead{(4)}    &
   \colhead{(5)}    &
   \colhead{(6)}    &
   \colhead{(7)}    &
   \colhead{(8)}    &
   \colhead{(9)}    &
   \colhead{(10)}   &
   \colhead{(11)}
%
   }
\startdata
\vspace{2.0ex} \\
C  & 2358$+$605 & J0001$+$6051 & 00 01 07.099852 & $+$60 51 22.79800 &  0.94 &  0.51  & $-$0.115  &  123  & $ 0.11 $ & $  0.11 $ \vspace{0.5ex} \\
C  & 2359$+$606 & J0002$+$6058 & 00 02 06.696680 & $+$60 58 29.83950 &  4.42 &  1.92  & $-$0.168  &   26  & $ 0.09 $ & $ <0.08 $ \vspace{0.5ex} \\
C  & 0002$+$541 & J0005$+$5428 & 00 05 04.363368 & $+$54 28 24.92414 &  0.60 &  0.47  & $ $0.167  &   79  & $ 0.34 $ & $  0.11 $ \vspace{0.5ex} \\
C  & 0003$+$505 & J0006$+$5050 & 00 06 08.249784 & $+$50 50 04.41150 &  0.68 &  0.81  & $-$0.041  &   64  & $ 0.16 $ & $  0.14 $ \vspace{0.5ex} \\
C  & 0005$+$568 & J0007$+$5706 & 00 07 48.468649 & $+$57 06 10.43705 &  1.81 &  2.18  & $ $0.631  &   46  & $ 0.08 $ & $  0.09 $ \vspace{0.5ex} \\
C  & 0012$+$610 & J0014$+$6117 & 00 14 48.792125 & $+$61 17 43.54198 &  0.57 &  0.29  & $-$0.034  &  153  & $ 0.25 $ & $  0.16 $ \vspace{0.5ex} \\
C  & 0024$+$597 & J0027$+$5958 & 00 27 03.286191 & $+$59 58 52.95899 &  0.56 &  0.34  & $-$0.214  &  136  & $ 0.23 $ & $  0.16 $ \vspace{0.5ex} \\
C  & 0032$+$612 & J0035$+$6130 & 00 35 25.310605 & $+$61 30 30.76122 &  0.94 &  0.52  & $-$0.038  &  117  & $ 0.13 $ & $  0.10 $ \vspace{0.5ex} \\
C  & 0034$-$220 & J0037$-$2145 & 00 37 14.825799 & $-$21 45 24.71171 &  1.17 &  2.78  & $-$0.834  &   59  & $ 0.09 $ & $  0.09 $ \vspace{0.5ex} \\
C  & 0039$+$568 & J0042$+$5708 & 00 42 19.451680 & $+$57 08 36.58569 &  0.39 &  0.25  & $ $0.046  &  162  & $ 0.48 $ & $  0.32 $ \vspace{0.5ex} \\
C  & 0041$+$677 & J0044$+$6803 & 00 44 50.759596 & $+$68 03 02.68540 &  0.67 &  0.29  & $-$0.163  &  154  & $ 0.23 $ & $  0.19 $ \vspace{0.5ex} \\
C  & 0044$+$566 & J0047$+$5657 & 00 47 00.428864 & $+$56 57 42.39373 &  0.53 &  0.39  & $ $0.006  &  154  & $ 0.18 $ & $  0.13 $ \vspace{1.0ex} \\
\vspace{1.5ex}
\enddata
\tablecomments{Table~\ref{t:cat} is presented in its entirety in the electronic
               edition of the Astronomical Journal. A portion is shown here
               for guidance regarding its form and contents. Units of right
               ascension are hours, minutes and seconds. Units of declination
               are degrees, minutes and seconds.
}
\ifpre
  \end{deluxetable*}
\else
  \end{deluxetable}
\fi

Total errors of our measurements of correlated flux density values for
sources stronger than $\sim$200~mJy were dominated by the accuracy of
amplitude calibration described above. This considers the error of
amplitude calibration as not exceeding 15\,\% and this estimate is
confirmed by our comparison of the flux densities integrated over the VLBA
images with the single-dish flux densities which we measured with
\mbox{RATAN-600} in 2006 for slowly varying sources without extended
structure. Details of the RATAN-600 single-dish observing program
including the method of observations and data processing can be found
in \citet{r:Kovalev_etal99,r:Kovalev_etal2002}.

\section{The catalogue of source positions}
\label{s:catalogue}

\begin{figure}[b!]
   \includegraphics[width=0.48\textwidth,clip]{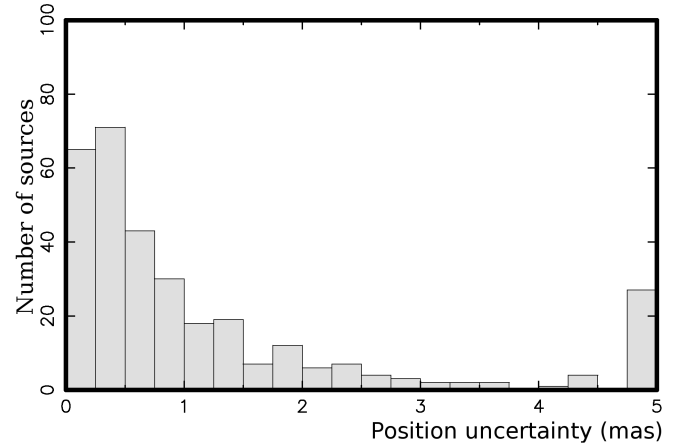}
   \caption{The histogram of the semi-major axes of inflated position
            error ellipses among 327 target sources in the VGaPS
            catalogue. \note{The last bin shows errors
            exceeding 4.75~mas.}
            }
   \label{f:pos_err}
\end{figure}

\Note{Of 487 sources observed, three or more detections were found for 327
objects.} After careful identification and removal outliers due to the
incorrect selection of the global maximum for weak sources with
\Note{SNR $ < 6$}, we selected 33,452 observations out of 59,690 from
three VGaPS experiments for analyzing in the single LSQ solution together
with 7.56 million other VLBI observations. The semi-major error ellipses of
inflated \Note{position errors for all sources except \object{0903+154} vary}
in the range 0.21 to 20~mas with the median value of 0.85~mas.
The histogram of position errors is shown in Figure~\ref{f:pos_err}.

The VGaPS catalogue is listed in Table~\ref{t:cat}.  Although
positions of all 5047 astrometric sources were adjusted in the LSQ
solution that included the VGaPS sources, only coordinates of 327
target sources observed in the VGaPS campaign are presented in the
table. The first column gives calibrator class: ``C'' if the source is
recommended as a calibrator or ``U'' if it has an unreliable position,
since there were less than 5~detections and there is a risk that the
secondary maximum of the coherent sum of weighted complex
cross-correlation samples has been picked and has not been flagged out.
The second and third columns give the IVS source name (B1950 notation)
and IAU name (J2000 notation). The fourth and fifth columns give
source coordinates at the J2000.0 epoch.  Columns /6/ and /7/ give
source position uncertainties in right ascension and declination in
mas after applying inflated errors according to \Note{equation~\ref{e:e31}}
(without
$\cos\delta$ factor), and column /8/ gives the correlation coefficient
between the errors in right ascension and declination. The number of
group delays used for position determination is listed in column
/9/. Column /10/ gives the estimate of the flux density integrated
over the entire map.  This estimate is computed as a sum of all CLEAN
components if a CLEAN image was produced. If we did not have enough
detections of the visibility function to produce a reliable image, the
integrated flux density is estimated as the median of the correlated
flux density measured at projected spacings less than
70~M$\lambda$. The integrated flux density is the source total flux
density with spatial frequencies less than 12~M$\lambda$ filtered out,
or in other words, the flux density from all details of a source with
size less than 20~mas. Column /11/ gives the flux density of
unresolved components estimated as the median of correlated flux
density values measured at projected spacings greater than
400~M$\lambda$. For some sources no estimates of the unresolved flux
density are presented, because either no data were collected at the
baselines used in calculations, or these data were unreliable.

  The on-line version of this catalogue is posted
on the Web\footnote{\tt \url{http://astrogeo.org/vgaps}}. For each
source it has 4 references: to a FITS file with CLEAN components of
naturally weighted source images; to a FITS file with calibrated
visibility data; to a postscript map of a source; and to a plot
of correlated flux density as a function of the length of the
baseline projection to the source plane.

\Note{In Table~\ref{t:not_det} we present a~priori coordinates,
total flux densities extrapolated to 24~GHz and spectral index estimates
for 160 target objects that have not been detected in the VGaPS experiment.
Some of these sources were detected in other VLBI astrometry
experiments at S/X bands.}

\begin{table}[tb!]
   \caption{\footnotesize
            \note{The list of 160 sources that have not been detected in
            VGaPS observations.}}
   \label{t:not_det}
   \centering
   \scriptsize
   \begin{tabular}{ l @{\quad} l @{\quad} l @{\quad} l @{\,} r  @{\quad} r  @{\enskip} r  @{\,} r  }
      \hline\hline
      \nntab{c}{Source names} & Right ascens. & Declination & gal. lat. & Flux & Sp.Ind. & \#       \\
      \ntab{c}{(1)} & \ntab{c}{(2)} & \ntab{c}{(3)} & \ntab{c}{(4)} &
      \ntab{c}{(5)} & \ntab{c}{(6)} & \ntab{c}{(7)} & (8)                                           \\
                        & & \ha h \hm m \hm s & \hp\ha $\degr \ha\ha' \ha\ha'' $  & deg & mJy & &   \\
      \hline
      2359$+$548 & J0002$+$5510 & 00 02 00.470 & $+$55 10 38.00 & $ -6.8$  &   121. & $-0.03$ &   2 \\
      0003$+$669 & J0006$+$6714 & 00 06 10.000 & $+$67 14 38.30 & $  5.0$  &   \n   & \n      &   0 \\
      0009$+$655 & J0012$+$6551 & 00 12 37.671 & $+$65 51 10.82 & $  3.5$  &   195. & $-0.59$ &   8 \\
      0010$+$722 & J0013$+$7231 & 00 12 58.750 & $+$72 31 12.76 & $ 10.1$  &   501. & $ 0.02$ &   8 \\
      0017$+$590 & J0020$+$5917 & 00 20 24.550 & $+$59 17 30.50 & $ -3.1$  &   324. & $-0.02$ &   6 \\
      0018$-$194 & J0021$-$1910 & 00 21 09.370 & $-$19 10 21.30 & $-79.6$  &   \n   & \n      &   0 \\
      0028$+$592 & J0031$+$5929 & 00 31 03.120 & $+$59 29 45.30 & $ -3.0$  &  3855. & $ 1.52$ &   2 \\
      0041$+$660 & J0044$+$6618 & 00 44 41.300 & $+$66 18 42.00 & $  3.7$  &   119. & $-0.91$ &   7 \\
      0107$+$562 & J0110$+$5632 & 01 10 57.553 & $+$56 32 16.93 & $ -5.9$  &   267. & $-0.66$ &  10 \\
      0113$+$241 & J0116$+$2422 & 01 16 38.067 & $+$24 22 53.72 & $-37.8$  &   168. & $-0.03$ &   7 \\
      0128$+$554 & J0131$+$5545 & 01 31 13.860 & $+$55 45 13.20 & $ -6.4$  &   150. & $-0.06$ &   5 \\
      \hline
   \end{tabular}
   \tablecomments{Table~\ref{t:not_det} is presented in its entirety in the electronic
                  edition of the Astronomical Journal. A portion is shown here
                  for guidance regarding its form and contents.
            \ynote{Columns 1--2 show IAU B1950 and J2000 source
            names, columns 3--5 show a~priori sources positions at the
            J2000.0 epoch, column 6 shows extrapolated a~priori total flux density
            at 24 GHz, column 7 shows coarse estimate of the spectral
            index, and column 8 shows the number of measurements of flux
            density found in the CATS database that were used for evaluation of the
            flux spectral index and extrapolation the flux density.}
}
\end{table}

\section{Comparison of K- and X/S-band astrometric VLBI positions}
\label{s:KSXcomparison}

\begin{figure}[t!]
   \includegraphics[width=0.48\textwidth,clip]{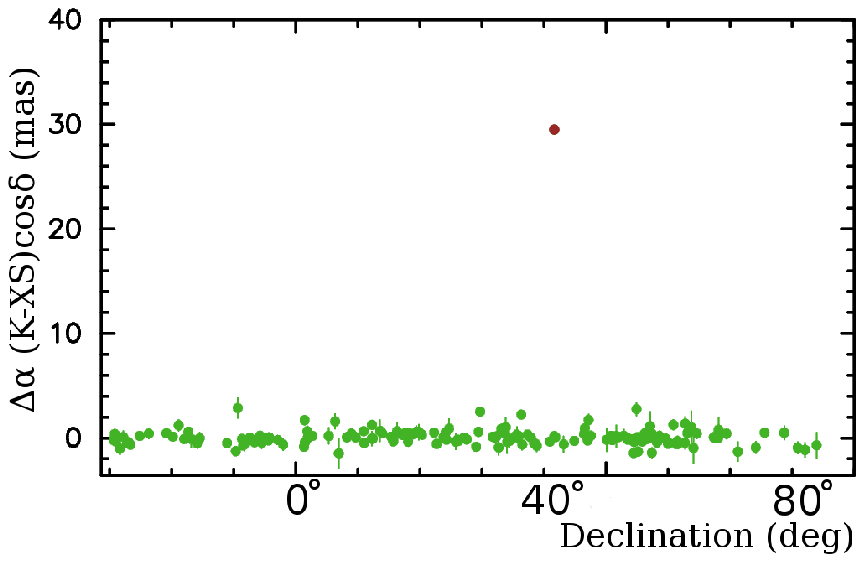}
   \par
   \includegraphics[width=0.48\textwidth,clip]{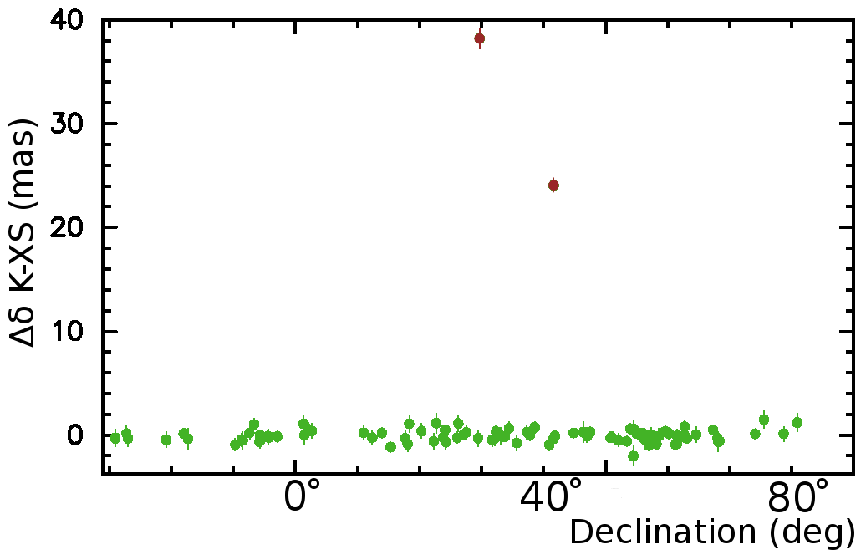}
   \caption{The differences in right ascensions scaled by $\cos \delta$
            (top) and declinations (bottom) from K~band VGaPS versus
            S/X historical VLBA observations among 164 sources with
            uncertainties less than 2~mas. Positions of two sources
            differ significantly.
   }
   \label{f:dif_kxs}
\end{figure}

  We searched the VLBI archive and found that among our target
sources, 206 were observed with S/X at the VLBA under VCS and
RDV programs before November 2010. We investigated the differences
in K-band observations against independent S/X observations.
We restricted our analysis to 192 objects that had uncertainties
from X/S and K~band solutions less than 5~mas.
Figures~\ref{f:dif_kxs} show the differences
in right ascensions and declinations.

\subsection{Special cases: 3C\,119 and 3C\,410}

\begin{figure}[t!]
   \centering
   \includegraphics[width=0.48\textwidth,trim=0cm 3cm 2cm 5.5cm,clip]{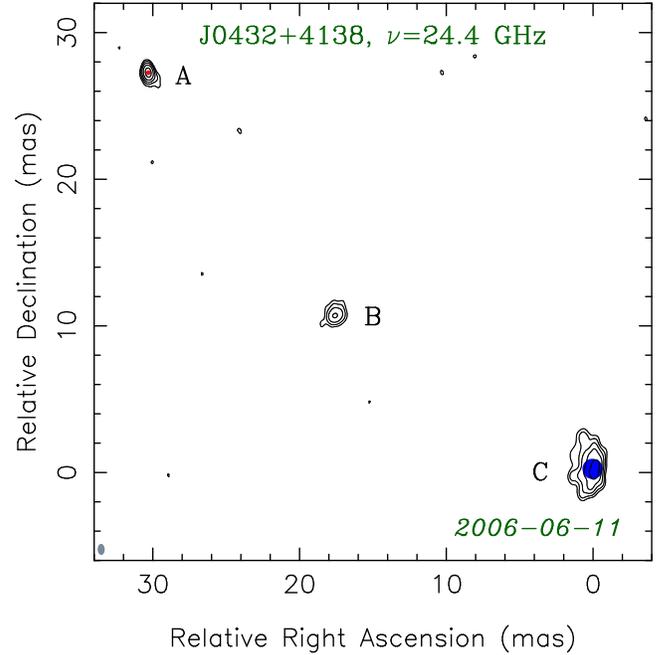}
   \caption{
      Naturally weighted K-band VGaPS CLEAN image of 3C119.
      The lowest contour of 4.1 mJy/beam
      is chosen at three times the rms noise,
      the peak brightness is 97 mJy/beam. The contour levels
      increase by factors of two. The dashed contours indicate
      negative brightness.
      The beam's full width at half maximum (FWHM) is shown
      in the bottom left corner of the images in grey.
      Red and blue spots indicate the positions and sizes (FWHM)
      of circular Gaussian model components for the features `A' and `C'
      respectively.
   }
   \label{f:J0432-K}
\end{figure}

\begin{figure}[t]
   \centering
   \includegraphics[width=0.44\textwidth,trim=4.4cm 0.3cm 5.5cm 2cm,clip]{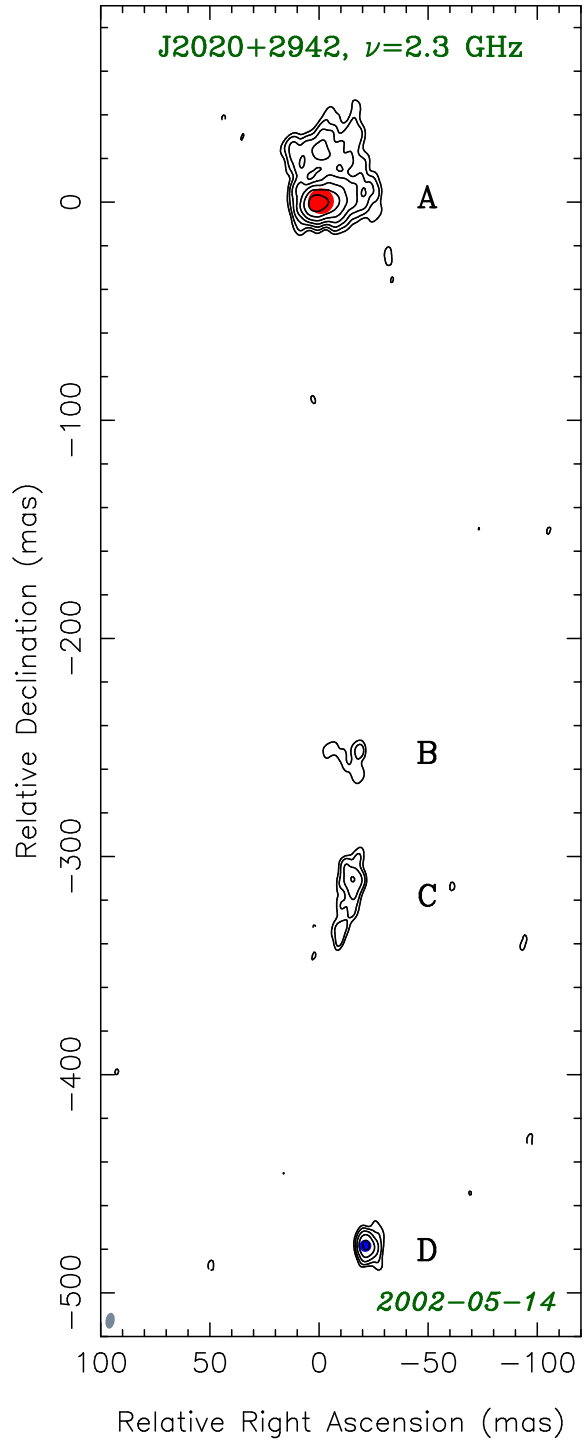}
   \caption{
     Naturally weighted S-band CLEAN image of 3C410 redone by us using
     VCS2 VLBA observations on 2002-05-14. The lowest contour of
     2.5 mJy/beam is chosen at three times the rms noise, the peak
     brightness is 242 mJy/beam. The contour levels increase by factors
     of two. The dashed contours indicate negative brightness. The
     beam's FWHM is shown in the bottom left corner of the images in grey.
     Red and blue spots indicate the positions and sizes (FWHM)
     of circular Gaussian model components for the features
     `A' and `D' respectively. It should be noted that feature
     A is very extended and the single Gaussian component does
     not represent it well.
   }
   \label{f:J2020-S}
\end{figure}

Positions of two sources, \object{J0432+4138} (also known as
\object{3C119}) and \object{J2020+2942} (\object{3C410}) are found
to be significantly off, namely 38.3 and 38.0~mas respectively.
The significance of this offset, 55 and 72 \Note{times of inflated
uncertainties, is too high to be explained by known error sources.}
RATAN-600 observations have shown that both
of them continuously show steep radio spectra and are slowly variable.
On VLBI scales, they were found to have significantly extended structure
(e.g., Figures~\ref{f:J0432-K},\ref{f:J2020-S}).

  One of these compact steep spectrum radio sources, the object
\object{J0432+4138}, is well studied at parsec scales. Recent
high-dynamic range images at 5 and 8.4~GHz and full bibliography
on historic observations can be found in \citet{r:man10}. This source
shows several bright components at 24~GHz. Two of them, namely
components `A' and `C' (Figure~\ref{f:J0432-K}), are located
40.6~mas apart. The feature C is stronger, its total flux density
is 0.54~Jy, but more \Note{extended. The FWHM} of a circular Gaussian
component fitted to the feature is found to be 1.4~mas. The feature A
is dimmer, 0.13~Jy, but more compact~--- 0.3~mas. We note that here and
below in this section the total flux density reported for different source
features is calculated as a sum of all CLEAN components representing
the corresponding structure. The Gaussian components fit is being
performed in the visibility plane.

  It is evident that the VGaPS 24~GHz
observations referred the position of the {\it weaker} A component,
while the S/X observations referred the position of the C component.
It is counter-intuitive. Since the component C is resolved, its contribution
to fringe amplitude is small at long baseline projections. At short baseline
projections the component C dominates, at long baseline projections the
component A dominates. But since the partial derivative of group delay
with respect to source position is proportional to the baseline length
projection on the source tangential plane, the contribution of long baseline
dominates in estimate of source position. According to \citet{r:man10},
the total flux density of components A and C at 8.4~GHz in 2001 was 70~mJy
and 1121~mJy respectively. This large difference between components'
flux densities is also supported by the previous VCS1 observations
of this object at S/X-bands.
The component C is less resolved at 8.4~GHz and
it dominates over A even at long baselines. Therefore, coordinate
of a source at X-band are closer to the C component. We checked it directly,
by suppressing observations at baselines longer than 800~km in a K-band
trial solution. The position estimate became close to the X/S position.
It is worth to note that the difference in position at K- and X-bands are
$29.2 \pm 0.4$ mas in right ascension and $24.1 \pm 0.4$ mas in declination,
while the offset of the component A with respect to the component C at K-band
is slightly larger: $30.25 \pm 0.10$~mas and $27.05 \pm 0.15$~mas.
The K-band and X-band position is not exactly the position of components
A and C, since in both solutions the contribution of the second
component is small but not entirely negligible.

  The nature of the difference in positions of J2020+2942 is similar.
It was observed in VCS2 experiment on May 15, 2002 \citep{r:vcs2}.
It had 63 detections at S-band and 72 detections at X-band. Position
of this source at S-band from analysis of only S-band data with applying
the ionosphere contribution from the GPS TEC model shows a very large
offset of $0.''5$ with respect to the X-band position (refer to
Table~\ref{t:pos_J2020+2942}). The errors of the ionosphere contribution
at S-band during the solar maximum affected position estimates
considerably, however not to that extent. Comparison of positions
of 130 sources with $\delta > 0$ from \Note{the solution that used only
S-band group delay observables} with respect to the X/S solution in that
experiment showed the differences in the range 2--7~mas. It is remarkable
that the X/S position is away from {\it both} X and S-band positions,
although intuitively we can expect them to be {\it between}
X and S positions. \Note{This
can be explained} if to surmise that the source J2020+2942 has two
components, $0''.48$ apart, one of them visible at X-band, but not
visible in the original S-band image, and another visible at S-band,
but not at X-band. An ionosphere-free linear combination of X and S band
observables is used in the X/S solution: $(1+\beta)\,\tau_{gx} - \beta\,\tau_{gs}$,
where $\beta = 1/\bigl(\omega^2_x/\omega^2_s-1\bigr)$ as it follows from
eqn.~\ref{e:e27}. In the case if position at S-band, $\vec{k}_s$,
is shifted with respect to the X-band position vector, $\vec{k}_x$,
the ionosphere free linear combination can be written as
\beq
    \tau_{if} = (1+\beta) \, \tau_{gx}(\vec{k}_x) \: - \:
                \beta \, \tau_{gs}(\vec{k}_x) \enskip - \enskip
                \beta \, \der{\tau}{\vec{k}} \, (\vec{k}_s - \vec{k}_x).
\eeq{e:e32}

  The first two terms correspond to the case of no offset between X and
S band positions. Therefore, estimates of the source position from the
ionosphere-free linear combination of observables will be shifted at
$-\beta$ with respect to the offset $(\vec{k}_s - \vec{k}_x)$, i.e. in
the {\it opposite} direction. Parameter $\beta$ was 0.076153 in the VCS2
experiment. Therefore, if our hypothesis that the X-band and S-band
observations detected emission from two components is true, then the shift
of X/S position with respect to the X-band position should be $-1.96$ mas
in right ascension and $-36.46$ \Note{in declination, just} within 0.3~mas
from reported X/S positions! The K-band position is within $1.5 \sigma$
from the X-band position.

\begin{table}[tb!]
   \caption{Position difference of J2020+2942 at different frequencies
            with respect to its position at X-band}
   \label{t:pos_J2020+2942}
   \centering
   \begin{tabular}{ l @{\qquad} r@{$\:$} r @{\qquad} r @{$\:$} r}
      \hline\hline
      Band & \nntab{c}{RA shift} & \nntab{c}{DEC shift} \\
      \hline
      X    & $   0.0  \: \pm $ & 0.58 \enskip mas & $    0.0 \: \pm $ & 0.71 \enskip  mas \\
      S    & $  25.77 \: \pm $ & 4.11 \enskip mas & $ 478.79 \: \pm $ & 4.39 \enskip mas \\
      X/S  & $  -2.29 \: \pm $ & 0.54 \enskip mas & $ -36.78 \: \pm $ & 0.66 \enskip mas \\
      K    & $   0.25 \: \pm $ & 0.41 \enskip mas & $   1.45 \: \pm $ & 0.70 \enskip mas \\
      \hline
   \end{tabular}
   \tablecomments{The offset in right ascension is scaled by $\cos\delta$.}
\end{table}

\begin{figure}[t]
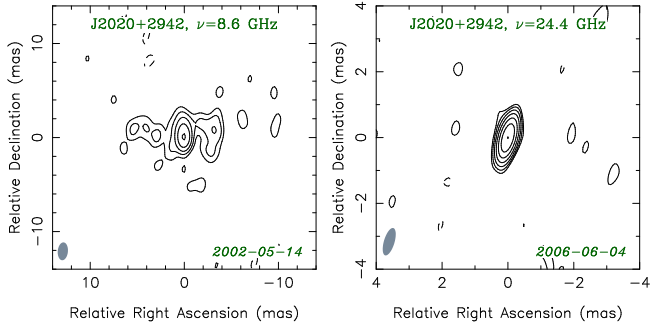

   \centering
\centering
\resizebox{1.0\hsize}{!}
{
  \includegraphics[trim=0cm 3cm 2cm 5cm,clip]{J2020+2942_X_2002_05_14_yyk.eps}
  \includegraphics[trim=0cm 3cm 2cm 5cm,clip]{J2020+2942_K_2006_06_04_yyk.eps}
}
\caption{
Naturally weighted X-band (\textit{left}, VCS2 data) and K-band
(\textit{right}, this survey --- VGaPS)
CLEAN images of 3C410. The lowest contour at X- and K-band is 3.3 and 2.8~mJy/beam
while the peak brightness is 56 and 178~mJy/beam respectively.
On the basis of our analysis, we identify the feature presented in this Figure
as feature `D' from Figure~\ref{f:J2020-S}.
   }
   \label{f:J2020-XK}
\end{figure}

  In order to check our hypothesis, we have re-imaged VCS2 observations
of J2020+2942 in a wide field and have detected several
previously unknown features `B', `C', `D', on a distance up to about
500~mas from the dominating extended structure `A' at S-band
(Figure~\ref{f:J2020-S}). The total flux density of the features
A and D is 1.40~Jy and 0.09~Jy respectively. The feature A is
significantly more extended then D. We have fitted two circular
Gaussian components to the $uv$-data in order to determine positions
of the features A and D. We note that the accuracy of position
determination for the component A from the image is very poor since
its structure being extended over at least 40~mas
is not well represented by a single Gaussian component.
The distance between Gaussian components A and D is 479~mas
while positional difference in right ascension and declination
is 25 and 479~mas respectively --- in a very good agreement with
independent astrometric measurements (Table~\ref{t:pos_J2020+2942}).

X-band and K-band wide field imaging (Figure~\ref{f:J2020-XK}) did not reveal
components with wide separation. We conclude from the astrometric analysis
presented above that X- and K-band images represent the more compact
component D. In this case we could also analyze its spectrum on the basis of
simultaneous S/X-band observations. Its total flux density at X-band is found
to be 0.25~Jy which provides the 2.3--8.6~GHz spectral index estimate
$\alpha=+0.8$ (flux density $\propto \nu^\alpha$) --- an indication of
synchrotron emission with significant self-absorption.
The features A, B, and C become too weak and/or too resolved that
we could detect then in the snapshot VLBA images with a limited
dynamic range and $uv$-coverage.

\subsection{General comparison: uncertainties, systematic K-S/X-band difference, and the core-shift effect}

\begin{figure}[tb!]
   \includegraphics[width=0.48\textwidth,clip]{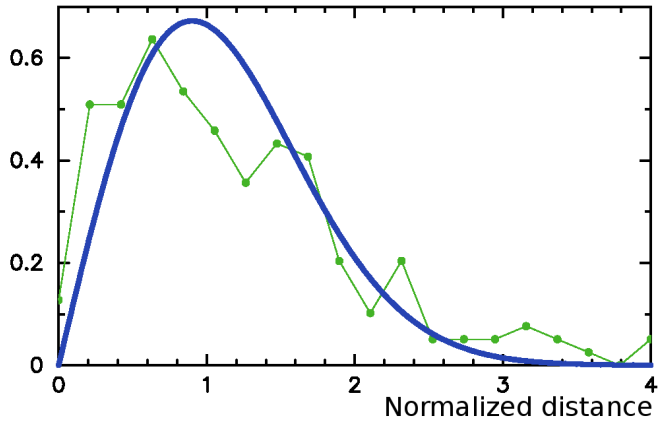}
   \caption{The empirical distribution of 190 normalized distances between
            K band position of target sources and their X/S positions
            (broken line) and the best fit Rayleigh distribution with
            $\sigma=0.90$ (solid thick line).
   }
   \label{f:nor_dist}
\end{figure}

   Position differences for other objects do not show peculiarities.
For instance, no declination dependent systematic differences similar to
those reported by \citet{r:kq} were found. The wrms of the differences is
0.46~mas in right ascension scaled by $\cos \delta$ and 0.61~mas in
declination. We have computed the normalized distances by dividing
them by $\sqrt{e_{k}^2 + e_{xs}^2}$, where $e_k$ is the projection of the
error ellipse of the K-band position to the direction of position
difference and $e_{xs}$ is the similar projection of the error ellipse of
the X/S position. In the case if position errors from K-band
and X/S catalogues are independent and Gaussian with the variance equal
to reported uncertainties, the distribution of normalized distances
will be Rayleigh with $\sigma=1$. The average of the normalized
distances over all sources, except \object{J0432+4138} and \object{J2020+2942},
is 1.276, only 2\% greater than the mean of the Rayleigh distribution,
$\sqrt{\pi/2}$. However, a close examination of the distribution
(see Figure~\ref{f:nor_dist}) reveals a slight deviation of its shape from
the shape of the Rayleigh distribution. The Rayleigh distribution that best
fits the distribution of normalized distances has $\sigma=0.90$. This is
an indication of a deviation of parent distributions from Gaussian.

   We can make several conclusions from this test. First, {\it on average}
reported formal uncertainties are correct within several percents.
Second, the effect of the core-shift is too small to contribute significantly
on results of single-epoch surveys. \YNote{According to \citet{r:kov08},
the typical apparent core-shift is expected to be 0.4~mas between
S- and X-bands. \citet{r:por09} stressed that when the core-shift
is proportional to $f^{-1}$, a source position derived from
ionosphere-free linear combinations of X and S-band group delays is
not sensitive to the core-shift and corresponds to a true position of
the jet base. The $f^{-1}$ core-shift dependence is expected for a conical
jet with synchrotron self-absorption in the regime of equipartition
between jet particle and magnetic field energy densities \citep{r:lob98}.
If we assume that this is indeed the case for majority of the sources
\citep[see also][]{r:sokol11}, the average
core-shift between K-band and effective S/X positions is reduced
to $0.4 \cdot \frac{f_\mathrm{x} f_\mathrm{s}}{f_\mathrm{k}\,
(f_\mathrm{x} - f_\mathrm{s})} = 0.06$~mas.}
Our observations would allow detecting the core-shift \Note{between
positions from S/X and K-bands observables} the 95\% confidence level
of a sample of 190 objects if the variance of the core-shift has been
greater than 1~mas. VGaPS observations set the upper limit of this
variance to 1~mas, which \Note{does not contradict to results
of core-shift measurements and predictions}.

\section{Summary}
\label{s:summary}

  In the VLBA Galactic Plane Survey we detected 327 compact radio sources
not previously observed with VLBI at 24~GHz in absolute astrometry mode.
Half of them are within $5^\circ$ of the Galactic plane; 206 of them were
also observed and detected within the VCS or RDV programs at S/X bands
in absolute astrometry mode. We determined K-band positions of all detected
sources. The position uncertainties \Note{for all but one source are in the
range} from 0.2 to 20~mas with the median value of 0.9~mas.
The quoted uncertainties account for various systematic effects and their
validity within several \Note{percents} was confirmed by comparison with
independent X/S observations. The detection limit of our observations was
in the range of 70--80~mJy. For the majority of detected sources parsec-scale
images were produced and correlated parsec-scale flux density estimated.
These results are presented in the form of the position catalog, calibrated
image and visibility data in FITS format, and visual plots.

The new wide-band fringe search baseline-oriented algorithm for
processing correlator output has been developed and implemented in the
software \PIMA.  It has reduces the detection limit of observations by
a factor of $\sqrt{N}$, where $N$ is the number of IF's, by
\Note{determining group delays, fringe phases at the reference frequency,
and phase delay rates from the coherent sum of the data from all IF's.}
The new algorithm increased the number of detected target sources by
a factor of 2.4 for this survey of weak \Note{objects} near the Galactic
plane. We validated the new algorithm by parallel processing of 1080 hours,
over 0.6 million observations, using both the traditional AIPS
approach and the new approach. The differences between source position
estimates processed with the wide-field and with the traditional AIPS
algorithms do not exceed 0.15~mas, which is satisfactory for any
practical application.

We investigated possible systematic errors caused by errors in the
ionosphere path delay derived from GPS TEC maps. We derived an
empirical model of the ionosphere driven delay path errors. We found
that for declinations $> -20^\circ$ for 90\% of the sources, mismodeling
path delay caused source position errors of less than 0.05~mas.  At
declinations below $-20^\circ$ these errors grow to 0.15~mas and for
some sources may reach 0.5~mas.

Comparison of new K~band VLBI positions with positions of 192 sources
observed at S/X showed an agreement with the wrms of 0.46 and 0.6~mas
in right ascensions and declination respectively, within reported position
uncertainties for all but two compact steep spectrum sources
\object{J0432+4138} and \object{J2020+2942}. For
these two objects, positional differences are about 40~mas. We showed
that the reason of these differences is that for sources with complex
extended structure positions were referred to different source details.
These two objects demonstrate an existence of an overlooked source
of errors in VLBI position catalogues
that will be studied in detail in the future.
A 1~mas upper limit on an apparent core-shift effect between 8 and 24~GHz
is found for the studied sample, in agreement with core-shift measurements
and predictions by \citet{r:kov08,r:sokol11,r:por09}.

\section{Acknowledgments}
\label{s:acknowledgments}

We would like to thank Leonid Kogan
for fruitful discussions about hidden secrets of correlators.
We used in our work the dataset MAI6NPANA provided by the
NASA/Global Modeling and Assimilation Office (GMAO) in the framework of
the MERRA atmospheric reanalysis project.
This project was started when YYK was working as a Jansky Fellow
of the National Radio Astronomy Observatory in Green Bank.
YYK was supported in part by the JSPS Invitation Fellowship
for Research in Japan (S-09143) and Russian Foundation
for Basic Research (08-02-00545 and 11-02-00368).
The National Radio Astronomy Observatory is a facility
of the National Science Foundation operated under cooperative
agreement by Associated Universities, Inc.
The authors made use of the database CATS of the Special
Astrophysical Observatory \citep{r:cats}.
This research has made use of the NASA/IPAC
Extragalactic Database \citep[NED,][]{r:ned} which is operated
by the Jet Propulsion Laboratory, California Institute of
Technology, under contract with the NASA.

{\it Facilities:} \facility{VLBA (project code BP125)}, \facility{RATAN-600}.

\end{document}